\documentclass{jpsj3}
\usepackage{color}
\def\ggs{\buildrel\textstyle > \over {\hbox{\raise0.2ex\hbox{$\sim$}}}}
\def\lls{\buildrel\textstyle < \over {\hbox{\raise0.2ex\hbox{$\sim$}}}}
\def\gsim{\, \lower0.75ex\hbox{$\ggs$}\, }
\def\lsim{\, \lower0.75ex\hbox{$\lls$}\, }
\newcommand{\beq}{\begin{equation}}
\newcommand{\eeq}{\end{equation}}

\newcommand{\beqa}{\begin{eqnarray}}
\newcommand{\eeqa}{\end{eqnarray}}
\newcommand{\non}{\nonumber}
\newcommand{\ii}{{\rm i}}

\def \Im{{\rm Im}}	
\def \Re{{\rm Re}}	
\def \nun{{\nu_n}}	
\def \epsilonn{{\epsilon_n}}	
\def \omegan{{\omega_n}}	
\def \e{{\rm e}}	
\def \Tr{{\rm Tr}}	

\def \Hl{H_{\lambda}}
\def \H0{H^{(0)}}

\def \ef{\varepsilon_f}	
\def \ep{\varepsilon}	
\def \Q{\hat Q}		

\def \phi{{\varphi}}
\def \dd{ {\rm d}}

\title
{
Non-Fermi Liquid and Fermi Liquid in Two-Channel Anderson Lattice Model:
Theory for Pr$A_2$Al$_{20}$ ($A$=V, Ti) and PrIr$_2$Zn$_{20}$
}

\author
{ Atsushi Tsuruta$^1$ and Kazumasa Miyake$^{1,2}$
}

\inst
{
$^1$Division of Materials Physics, Department of Materials Engineering Science,
Graduate School of Engineering Science, Osaka University, Toyonaka, Osaka
560-8531, Japan\\
$^{2}$Toyota Physical and Chemical Research Institute, Nagakute, Aichi 480-1192, Japan\\
}

\recdate
{
May 28, 2015
}

\abst
{
We theoretically investigate electronic states and physical properties in a two-channel Anderson lattice model to understand the non-Fermi liquid behaviors observed in PrV$_2$Al$_{20}$ and PrIr$_2$Zn$_{20}$, whose ground state of the crystalline electric field for a local $f$-electron is the $\Gamma_3$ non-Kramers doublet of $f^2$-configuration and whose excited state is the $\Gamma_7$ Kramers doublet of $f^1$-configuration.
We use the expansion from the limit of the large degeneracy $N$ of the ground state ($1/N$-expansion), with $N$ being the spin-orbital degeneracy.
The inclusion of the self-energy of conduction electrons up to the order of $O(1/N)$ leads to heavy electrons with channel and spin-orbit degeneracies.
We find that the electrical resistivity is proportional to the temperature $T$ in the limit $T\to0$ and follows the $\sqrt{T}$-law in a wide temperature region, i.e., $T_x<T<T_0$, where the typical values of $T_x$ and $T_0$ are $T_x\sim10^{-3}T_{\rm K}$ and $T_0\sim10^{-2}T_{\rm K}$, respectively, $T_{\rm K}$ being the Kondo temperature of the model.
We also find non-Fermi liquid behaviors at $T\ll T_{\rm K}$ in a series of physical quantities; chemical potential, specific heat, and magnetic susceptibility,
which explain the non-Fermi liquid behaviors observed in PrV$_2$Al$_{20}$ and PrIr$_2$Zn$_{20}$.
At the same time, we find that the Fermi liquid behavior becomes prominent for the system with a small hybridization between $f$- and conduction electrons, explaining the Fermi liquid behaviors observed in PrTi$_2$Al$_{20}$.
}

\begin{document}

\sloppy
\maketitle
\section{Introduction}

\begin{figure}
\includegraphics[width=10cm]{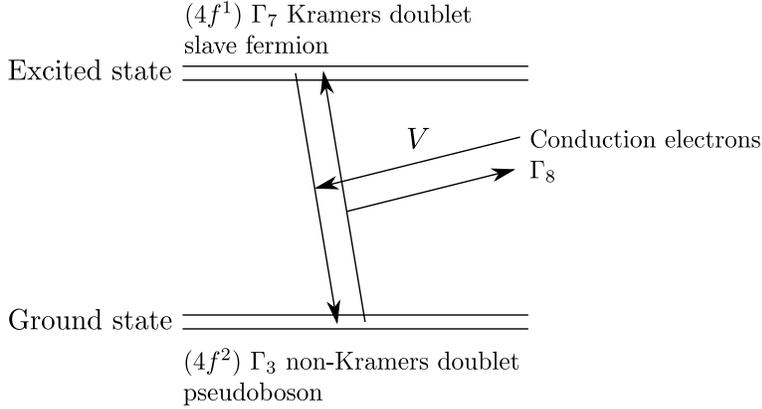}
\caption{Levels scheme of $f$-electrons and hybridization path with conduction electrons of model Hamiltonian, Eqs. (2)--(5).}
\label{Fig:level}
\end{figure}

Recently, non-Fermi liquid behaviors in the $T$ dependence of resistivity, $\rho(T)$, have been reported in PrV$_2$Al$_{20}$\cite{Sakai} and PrIr$_2$Zn$_{20}$\cite{Onimaru}.
Namely, the electrical resistivity is proportional to $\sqrt{T}$
in a wide temperature region.
The $T$ dependence of specific heat, $C(T)$, and $T$ dependence of magnetic susceptibility, $\chi_m(T)$, increase in proportion to $({\rm const.}-\sqrt{T})$ toward $T_Q$, the transition temperature of quadrupolar ordering, as $T$ decreases below the Kondo temperature $T_{\rm K}$, which is a fundamental energy scale characterizing the physics. 
The anomaly in the specific heat and the cusp in the magnetic susceptibility are also observed in PrPb$_3$\cite{Sato}.
From the analyses of the specific heat, magnetic moment, and inelastic neutron scattering experiment, the ground state of the crystalline-electrical field (CEF) of the local $f$-electron is considered to be the $\Gamma_3$ non-Kramers doublet in $4f^2$-configuration\cite{Onimaru}, as shown in Fig. \ref{Fig:level}.

Such a system with the $\Gamma_3$ CEF ground state in $f^2$-configuration is expected to exhibit anomalous behaviors associated with the two-channel Kondo effect.
Indeed, the two-channel Kondo model (proposed by Nozi\`eres and Blandin\cite{Nozieres}) was revived in late 1980s when Cox proposed it as a realistic model for explaining the anomalous non-Fermi liquid properties of the cubic heavy fermion metal UBe$_{13}$ on the basis of the quadrupolar Kondo effect.\cite{Cox1}
Since then, the two-channel Kondo problem has been attracting much attention.\cite{Cox2}

On the other hand, a strong electronic correlation has been widely considered to play a crucial role in the generation of the heavy electrons observed in rare-earth and actinide-based metals as well as the various anomalous phenomena observed in high-$T_{\rm c}$ cuprate superconductors.
For the simulation of such systems with a strong electronic correlation, various Anderson models have been used. 
One of the powerful methods to properly treat these models is that based on the expansion from the limit of the large spin-orbital degeneracy $N$ ($1/N$-expansion method).

In the case of the multichannel impurity Anderson model,
Cox et al.\cite{Cox2,Cox3} showed that the spectral weights of the slave boson and the pseudofermion diverge in proportion to $\omega^{-\alpha}\theta(\omega)$, near $\omega\sim0$, with  $\alpha=N/(N+M)$ and $M/(N+M)$, respectively, where $M$ is the degeneracy of the channel,
with the use of the noncrossing approximation (NCA), which is valid in the region of $N,M\gg1$, regardless of $n_f$, 
where $n_f$ is the average number of localized electrons per channel at the impurity site.
On the other hand, Tsuruta et al.\cite{Tsuruta2} showed with the use of $1/N$-expansion 
that these exponents are given by $\alpha=1-n_f^2M/N$ and $(2n_f-n_f^2)M/N$, respectively.
These exponents $\alpha$'s are valid in the region of $N\gg 1$, and either $N\gg M$ or $M=1$.
In the limit of $n_f\to 1$, these exponents of the pseudoparticle spectra and physical quantities, as shown above, are in agreement with those obtained by NCA in the limit of $N\gg M$.

\=Ono et al.\cite{Ono} have treated the conventional (single-channel) 
Anderson lattice model by the $1/N$-expansion method in the leading order of $1/N$, 
and have succeeded in describing the heavy Fermi liquid state in which 
the Luttinger sum rule holds.  
Various Fermi liquid properties of the lattice model have been discussed 
along this formulation by taking into account higher-order terms in $1/N$.~\cite{Ono2,Ono3}
Tsuruta et al.\cite{Tsuruta4,Tsuruta5} have also confirmed, 
in the single-channel Anderson lattice model, that the imaginary part of the self-energy of 
conduction electrons is given by the form 
${\rm Im}\Sigma(\epsilon+i0_+)=-\alpha[\epsilon^2+\left(\pi T\right)^2]$, $\alpha$ being 
a constant proportional to $T_{\rm K}^{-2}$, up to $O(N^{-2})$ in the regions of 
frequency $|\epsilon|\ll T_{\rm K}$ and of temperature $T\ll T_{\rm K}$, and that 
the Luttinger sum rule holds up to $O(N^{-1})$.
Furthermore, it has been confirmed that 
${\rm Im}\Sigma(\epsilon+i0_+)\propto-\ln[\max(|\epsilon|,T)]$ at 
$\max(|\epsilon|,T)>T_{\rm K}$, implying that the Kondo effect at $T>T_{\rm K}$ 
is reproduced.~\cite{Tsuruta4,Tsuruta5}
Namely, 
this order of the $1/N$-expansion method is a powerful method for investigating the essential properties of the Anderson lattice model.

On this formalism,
Nishida et al.\cite{Nishida} have investigated the effects of the CEF level splitting on the resistivity of a Ce-based compound taking into account the self-energy up to $O(1/N)$,
and have explained the $T$ dependence of the resistivity in CeCu$_2$(Si,Ge)$_2$, which exhibits the double-peak structure in its $T$ dependence at ambient pressure.
They have also shown that the double peaks merge into a single peak as the $c$-$f$ 
hybridization increases, explaining such a mergence under pressure observed in
CeCu$_2$(Si,Ge)$_2$\cite{Jacard,Holmes} and some other heavy fermion systems, such as 
CeAl$_2$.~\cite{Barbara}

The dynamical mean field theory (DMFT) is exact in infinite dimensions ($d=\infty$) and explains the existence of the Mott transition in the Hubbard model,\cite{Georges}
and the Kondo insulator and heavy electron state in the Anderson lattice model.\cite{Jarrell3,Shimizu}
The properties of the two-channel Kondo lattice model in $d=\infty$
have been studied by the use of quantum Monte Carlo simulation\cite{Jarrell,Jarrell2,Anders} and the continuous-time quantum Monte Carlo method\cite{Hoshino} as solvers of the impurity problem, in which
the phase diagrams of superconductivity, antiferromagnetic or quadrupolar ordering have been found.
It has also been found that the resistivity $\rho(T)$ is finite even at $T=0$, 
which is apparently unphysical.
This unphysical result has occurred since the translational symmetry was not properly 
taken into account in their study, which relies on a property of $d=\infty$.
This is because, in $d=\infty$, the intersite correlation effects on the self-energy of the conduction electrons are inevitably neglected.

On the other hand, on the basis of the $1/N$-expansion formalism, Tsuruta et al.\cite{Tsuruta4} have shown, 
in the two-channel Anderson lattice model shown in Fig. 1, 
that the imaginary part of the self-energy of conduction electrons is proportional to $T$ 
in the limit of $T\to0$ up to $O(1/N)$, if the intersite effects on the self-energy are 
properly taken into account.  
Indeed, the imaginary part of the self-energy of conduction electrons, at $|\epsilon|\ll T_{\rm K}$ 
and $T\ll T_{\rm K}$, is given in the form
\beqa
{\rm Im}\Sigma_{\mib{k}}(\epsilon+i0_+)&=&
-\pi \alpha^{(1)}_{\mib{k}}[\epsilon^2+\left(\pi T\right)^2]
-\pi \frac{\beta^{(1)}_{\mib{k}}}{1+T^*_{\mib{k}}/T}(1-M^{-2}),
\eeqa
where
$T^*_{\mib{k}}$ is the temperature characterizing the non-Fermi liquid state,
and $\alpha^{(1)}_{\mib{k}}$ and $\beta^{(1)}_{\mib{k}}$ 
are constants proportional to $N_{\rm F}V^2T_{\rm K}^{-2}$ and $N_{\rm F}^{-1}$, where $V$ and $N_{\rm F}$ are $c$-$f$ hybridization and the density of states of conduction electrons at the Fermi level, respectively.
The imaginary part with $\epsilon=0$ is proportional to $T$ in the limit $T\to0$ in the 
leading order of $T$, i.e., it exhibits the non-Fermi-liquid-type behavior; furthermore, in a rather wide temperature region of $T<T_{\rm K}$, it behaves as 
${\rm Im}\Sigma_{\mib k}(i0_+)\propto\sqrt{TT_{\rm K}}$.
Thus, the $T$ and $\epsilon$ dependences of ${\rm Im}\Sigma_{\mib{k}}(\epsilon+i0_+)$ given by Eq. (1) 
may explain anomalous properties observed in 
PrV$_2$Al$_{20}$\cite{Sakai} and PrIr$_2$Zn$_{20}$\cite{Onimaru}.

In this paper, we follow the idea that these Pr-based compounds are the realization of the two-channel Anderson lattice model and investigate the non-Fermi liquid behavior of the two-channel Anderson lattice model by calculating the $T$ dependences of electrical resistivity, chemical potential, specific heat, and magnetic susceptibility in the low-temperature region $T<T_{\rm K}$.
In Sect. 2, we briefly review the discussions found in Ref. \citen{Tsuruta5}; i.e., 
we show the model and formulation, and explicitly calculate the vertices and the self-energy for conduction electrons.
In Sects. 3, 4, 6, and 7, we calculate the $T$ dependences of chemical potential [$\mu(T)$], electrical resistivity [$\rho(T)$], specific heat [$C(T)$], and magnetic susceptibility [$\chi_m(T)$], respectively.
In Sect. 5, we analyze the origin of the difficulty in using the DMFT when applied to the multichannel Anderson lattice model.
In Appendix A, we show the relationship among self-energies in the impurity model, $d=\infty$ lattice model (treated by DMFT), and the lattice model in finite dimensions (treated by $1/N$-expansion formalism).
In Appendix B, we show the formalism for calculating the specific heat in terms of the renormalized Green function of conduction electrons, pseudobosons, and slave fermions.
A recipe for numerical calculations is given in Appendix C.


\section{Review of the Previous Investigation of the Two-channel Anderson Lattice Model}
\subsection{Model and formal preliminaries}

As shown in Fig. 1, we introduce an Anderson lattice model for Pr1-2-20 compounds, whose CEF ground state of f-electrons is the $\Gamma_3$ non-Kramers doublet in $f^2$-configuration, as follows:
\beqa
&&H=H_c+H_f+H_v,\label{Eq:Horg}\\
&&H_c=\sum_{\sigma=1}^M\sum_{\tau_1,\tau_2=1}^{N}\sum_{\mib{k}}\ep_{\mib{k}\tau_1\tau_2}c^+_{\mib{k}\tau_1\sigma} c_{\mib{k}\tau_2\sigma},\label{Eq:Horg1}\\
&&H_f=\sum_{i}\sum_{\tau=1}^{N}\ep^{(0)}_{\Gamma_3}|f^2:i \tau\rangle\langle f^2:i \tau|
+\sum_{i}\sum_{\sigma=1}^{M}\ep^{(0)}_{\Gamma_7}|f^1:i \sigma\rangle\langle f^1:i \sigma|,\label{Eq:Horg2}\\
&&H_v=\frac{1}{\sqrt{N_L}}\sum_{\mib{k}i}\sum_{\tau\sigma}\left(Ve^{-i\mib{k}\cdot\mib{R}_i}c^+_{\mib{k}\tau{\bar \sigma}}|f^1:i \sigma\rangle\langle f^2:i \tau|
+{\rm h.c.}\right),\label{Eq:Horg3}
\eeqa
where $\sigma$ and $\tau$ denote the components of the spin-orbital degeneracy $M(=2)$ of the $\Gamma_7$ and $\Gamma_8$ CEF states in $f^1$-configuration and the quadrupole moment $N(=2)$ of the $\Gamma_3$ CEF state in $f^2$-configuration, respectively, and $\bar{\sigma}$ denotes the opposite component from $\sigma$;
$c_{\mib{k}\sigma\tau}$ represents the annihilation operator of a conduction electron with the wave vector $\mib{k}$, the spin-orbital $\sigma$, and the quadrupole moment $\tau$; $|f^2:i\tau\rangle$ is the $f^2$ state with $\tau$ at the $i$-site, and $|f^1:i\sigma\rangle$ is the $f^1$ state with $\sigma$ at the $i$-site; V represents the hybridization transforming from $|f^2:i\tau\rangle$ to $|f^1:i\sigma\rangle\otimes c_{\mib{k}\bar{\sigma}\tau}$ and vice versa (see Fig. 1); and $N_L$ is the total number of lattice sites. 
Hereafter, we call $\sigma$ the channel degrees of freedom and $\tau$ the pseudospin degrees of freedom.
With the use of pseudoparticles, we can rewrite the Hamiltonian given by Eq. (\ref{Eq:Horg}) as
\beqa
&&H=\sum_{\sigma=1}^M\sum_{\tau_1,\tau_2=1}^{N} \sum_{\mib{k}} \ep_{\mib{k}\tau_1\tau_2} c^+_{\mib{k}\tau_1\sigma} c_{\mib{k}\tau_2\sigma}
+\sum_i\sum_{\tau=1}^N \ep^{(0)}_{\Gamma_3} b^+_{i\tau} b_{i\tau}
+\sum_i\sum_{\sigma=1}^M \ep^{(0)}_{\Gamma_7} f^+_{i\sigma} f_{i\sigma}\non\\
&&\hspace{2em} +\frac{1}{\sqrt{N_L}}\sum_{\sigma=1}^M\sum_{\tau=1}^N\sum_{i, \mib{k}} \left (V c^+_{\mib{k}\tau\bar{\sigma}} b_{i \tau} f^+_{i\sigma}e^{-i \mib{k}\cdot\mib{R}_i} + {\rm h.c.} \right ),\non\\
&&\label{H}
\eeqa
where we have introduced the pseudoboson annihilation operator $b_{i\tau}$ for representing the $|f^2:i\tau\rangle$ state and the slave fermion annihilation operator $f_{i\sigma}$ for representing the $|f^1:i\sigma\rangle$ state.
Although we mainly investigate the case of two channels ($M=2$) in this study, the case of the single channel ($M=1$) is also discussed for comparison.
In the latter case, both the ground state of $f^1$-configuration and the conduction electrons are specified by a Kramers doublet with $\sigma$ and the excited state is in $f^0$-configuration, in contrast to the situation shown in Fig. 1.
Although $\ep_{\mib{k}\tau{\bar \tau}}\ne0$ in general, there occurs no qualitative difference from the case of $\ep_{\mib{k}\tau{\bar \tau}}=0$, where $\bar{\tau}$ denotes the opposite component from $\tau$.
Thus, in this paper, we restrict our discussion to the case of $\ep_{\mib{k}\tau{\bar \tau}}=0$.

To guarantee the equivalence between the transformed model [Eq. (\ref{H})] and the original model [Eq. (\ref{Eq:Horg})], the Hamiltonian [Eq. (\ref{H})] must be treated within the subspace where the local constraint
\beq
\Q_{i} = \sum_{\tau}b^+_{i \tau} b_{i \tau} + \sum_{\sigma} f^+_{i\sigma} f_{i\sigma} = 1, \label{Q}
\eeq
is fulfilled.
To calculate physical quantities within the subspace restricted by the local constraint [Eq. (\ref{Q})], we evaluate the expectation value $\langle\hat{A}\rangle$ of a physical quantity $\hat{A}$ such that\cite{Coleman,Jin}  
\beqa
\left\langle \hat{A} \right\rangle = \lim_{\{\lambda_i\} \rightarrow \infty}
		\left( {\left\langle \hat{A} \Pi_i \Q_{i} \right\rangle}_\lambda / 
		{\left\langle \Pi_i \Q_{i} \right\rangle}_\lambda \right), 
			 \label{expectation value}
\eeqa
where
\beqa
\langle \hat{A} \Pi_i \Q_{i} \rangle _\lambda \equiv
		\Tr[\e^{-\beta \Hl} \hat{A} \Pi_i \Q_{i}]/Z_\lambda,
\eeqa
 with
\beqa 
&&Z_\lambda \equiv \Tr[\e^{-\beta \Hl}],\\
&&\Hl \equiv H + \sum_i \lambda_i \Q_{i}.
\eeqa
In order to calculate Eq. (\ref{expectation value}) explicitly, we employ the perturbation expansion in the power of $1/N$ following the rules as
\beqa
\frac1{N_L}\sum_{\tau}\sum_{\mib{k}}1=O\left[ (1/N)^0  \right],
\eeqa
and
\beqa
\frac1{N_L}\sum_{\mib{k}}1=O\left( 1/N  \right).\label{O1N}
\eeqa
In Refs. \citen{Ono} and \citen{Nishida}, one can see the validity of this rule of power counting in $1/N$ and its physical meaning behind it.
For explicit calculation in this paper, we set $N=2$, which may not lose the generality because we do not use the condition $1/N\ll 1$ explicitly.

\subsection{Self-energy of $O[ (1/N)^0]$}

\begin{figure}
\includegraphics[width=10cm]{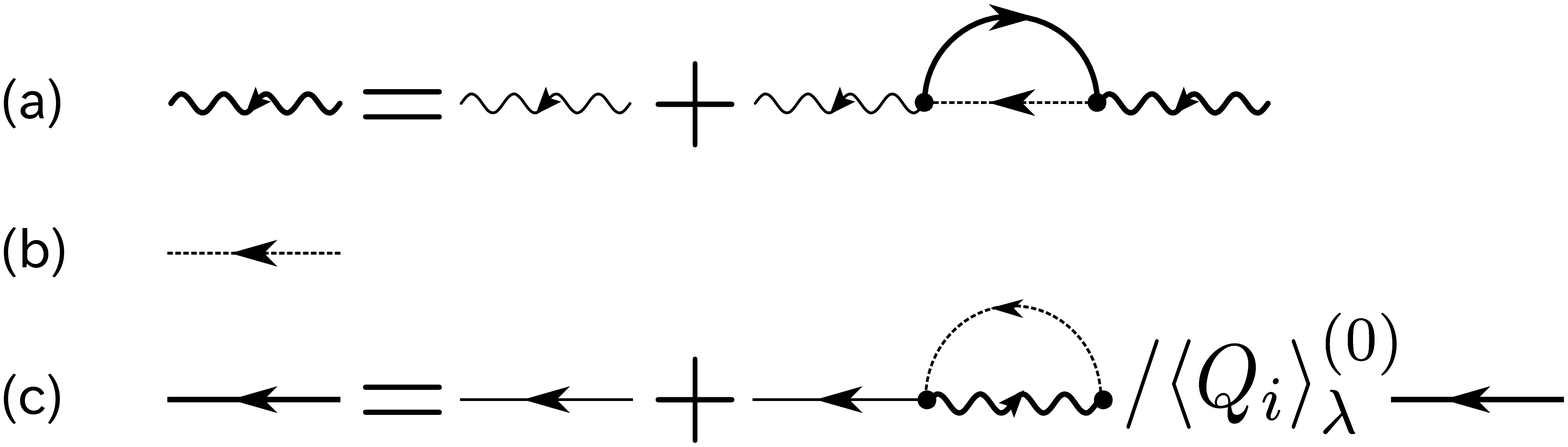}
\caption{Feynman diagram representation of the Dyson equations for the single-particle Green functions on the order of $O[(1/N)^0]$: (a) $F_{i\sigma}^{(0)}(i\epsilon_n)$ for the slave fermion, (b) $B_{i\tau}^{(0)}(i\nu_n)$ for the pseudoboson, and (c) $G_{\mib{k}\tau\sigma}^{(0)}(i\epsilon_n)$ for the conduction electron.}
\label{Fig:N0}
\end{figure}

By including terms on the order of $(1/N)^0$,
the Green functions for the slave fermion $F_{i\sigma}^{0}(i\epsilon_n)=(i\epsilonn-\lambda_i-\ep_{\Gamma_7})^{-1}$,
where $\ep_{\Gamma_7}=\ep^{(0)}_{\Gamma_7}-\mu$,
with $\mu$ being the chemical potential,
and conduction electrons 
$G_{\mib{k}\tau\sigma}^{0}(i\epsilon_n)=(i\epsilonn-\xi_{\mib{k}})^{-1}$,
where $\xi\equiv \ep_{\mib{k}}-\mu$, are renormalized by the effect of self-energies as illustrated in Fig.~\ref{Fig:N0},
while the Green function for the pseudoboson $B_{i\tau}^0(i\nu_n)=(i\nu_n-\lambda_i-\ep_{\Gamma_3})^{-1}$,
where $\ep_{\Gamma_3}=\ep^{(0)}_{\Gamma_3}-2\mu$, is unrenormalized.
In the limit of $T\to0$, the Green functions $F^{(0)}_{i\sigma}$, $B^{(0)}_{i\tau}$, and $G^{(0)}_{\mib{k}\tau\sigma}$ are given by 
\beqa
F^{(0)}_{i\sigma}(i \epsilonn ) &=& \frac{a}{i\epsilonn -\lambda_i -\left(\ep_{\Gamma_3} -E_0\right)}+C^{(0)}(i\epsilonn), \label{F0}\\
B^{(0)}_{i\tau} (i\nun) &=& \frac1{i\nun -\lambda_i -\ep_{\Gamma_3}}, \label{B0}\\
G^{(0)}_{\mib{k}\tau\sigma} (i\epsilonn) &=& \frac{1}{ i \epsilonn -\xi_{\mib{k}} -\Sigma^{(0)}_{\mib{k}\tau\sigma}(i\epsilonn)},\label{G0}
\eeqa
where $a$ is the residue of the slave fermion,
$E_0$ is the binding energy of the slave fermion relative to the chemical potential $\mu$ and corresponds to the Kondo temperature $T_{\rm K}=D\exp(-1/|J|N_{\rm F})$ in the impurity Anderson model, where $D$ is half the bandwidth of the conduction electrons, $J$ is the $c$-$f$ exchange interaction, and $N_{\rm F}$ is the density of states of the conduction electrons.
Hereafter, $E_0$ is used in theoretical expressions while $T_{\rm K}$ is used
in figures that can be compared with experiments.
The point is that $E_0$ and $T_{\rm K}$ are equivalent quantities.
Note that $(\ep_{\Gamma_3}-E_0)$ in Eq. (\ref{F0}) is the effective level of the $f^1$-state renormalized by the self-energy of $O[(1/N)^0]$.
$\Sigma_{\mib{k}}^{(0)}(i\epsilonn)$ in Eq. (\ref{G0}) is the self-energy of the conduction electrons of $O(1/N)$.
Note that $\epsilonn$ in Eqs. (\ref{F0}) and (\ref{G0}) is the fermionic Matsubara frequency, $\epsilonn=(2n+1)\pi T$,
and
$\nun$ in Eq. (\ref{B0}) is the bosonic Matsubara frequency, $\nun=2n\pi T$.
$C^{(0)}(i\epsilonn)$ in Eq. (\ref{F0}) represents the incoherent part of the slave fermion far from the Fermi level (or the chemical potential $\mu$), and is given by
\beqa
C^{(0)}(i\epsilonn)&=&a(i\epsilonn-\lambda-\ep_{\Gamma_3}+\ep_{\Gamma_7}+E_0)\frac{V^2}{N_L}\sum_{\mib{k},\sigma}F_{i\sigma}^{(0)}(i\epsilonn)\non\\
&&\times\int d\epsilon f(\epsilon)\frac{-1}{\pi}{\rm Im}G^{(0)}_{\mib{k}\tau\sigma}(\epsilon+i0_+)\frac{1}{(\epsilon-E_0)^2(i\epsilonn+\epsilon-\lambda-\ep_{\Gamma_3}+\ep_{\Gamma_7})},\non\\
\label{C0}
\eeqa
where $f(x)$ is the Fermi distribution function $f(x)\equiv 1/(e^{x/T}+1)$.
One can see from Eq. (\ref{C0}) that $\Im C(\epsilon+i0_+)$ is nonzero only in the region $\epsilon\ge\ep_{\Gamma_3}-\ep_{\Gamma_7}+\lambda_i$.
On the other hand, $\Sigma_{\mib{k}\tau\sigma}^{(0)}(i\epsilonn)$ in Eq. (\ref{G0}) is given by
\beqa
\Sigma^{(0)}_{\mib{k}\tau\sigma}(i\epsilonn) &\equiv& \frac{V^2}{N_L}\sum_i\lim_{\lambda_i\rightarrow\infty}\left[-T\sum_{\nu_n}F^{(0)}_{i\sigma}(i\epsilonn+i\nu_n)\right.\non\\
&&\hspace{8em}\times\left.B^{(0)}_{i\tau}(i\nu_n)/\langle \hat{Q}_{i}\rangle^{(0)}_{\lambda}\right],\label{Sigma01}
\eeqa
where $\langle \hat{Q}_i \rangle^{(0)}_{\lambda}$ denotes the contributions of $O[(1/N)^0]$ to $\langle \hat{Q}_i \rangle_{\lambda}$ in the $1/N$-expansion:
\beqa
&&\langle \hat{Q}_i \rangle^{(0)}_{\lambda}=M\e^{-\beta(\lambda_i+\ep_{\Gamma_3}-\ep_{\Gamma_7}-E_0)}.
\eeqa
Then, the contribution of ${\rm Im}C(\epsilon+i0_+)$ to $\Sigma_{\mib{k}\tau\sigma}^{(0)}(i\epsilonn)$ is smaller than the imaginary part by a factor of $\e^{-\beta E_0}\to0$ (as $T\to0$).
Therefore, $\Sigma_{\mib{k}\tau\sigma}^{(0)}(i\epsilonn)$ is approximately given by
\beqa
\Sigma_{\mib{k}\tau\sigma}^{(0)}(i\epsilonn)&\simeq&\frac{1}{M}\frac{aV^2}{i\epsilonn-E_0},\label{Sigma0}
\eeqa
where $E_0$ and $a$ are determined by solving the following coupled equations:
\beqa
 \ep_{\Gamma_3}-\ep_{\Gamma_7}-E_0 -\frac{V^2}{N_{\rm L}}\sum_{\mib{k},\tau}
 \int d\epsilon f(\epsilon)\frac{-1}{\pi}{\rm Im}G^{(0)}_{\mib{k}\tau\sigma}(\epsilon+i0_+)\frac{1}{\epsilon-E_0}=0,\label{Eq:E00}
\eeqa
and
\beqa
\frac1a= 1+\frac{V^2}{N_{\rm L}}\sum_{\mib{k},\tau}
 \int d\epsilon f(\epsilon)\frac{-1}{\pi}{\rm Im}G^{(0)}_{\mib{k}\tau\sigma}(\epsilon+i0_+)\frac{1}{(\epsilon-E_0)^2}.\label{Eq:a0}
\eeqa
The residue $a$ is approximately given by $a\simeq E_0/N_{\rm F}V^2$,
since the second term on the r.h.s. of Eq. (\ref{Eq:a0}) is approximately given by $N_{\rm F}V^2/E_0$ if we note that the approximate relation $\frac1{N_L}\sum_{\mib{k}}(-1/\pi){\rm Im}G^{(0)}_{\mib{k}\tau\sigma}(\epsilon+i0_+)\simeq N_{\rm F}$ holds.
With the use of this self-energy, the Green function $G^{(0)}_{\mib{k}\tau\sigma}(i\epsilonn)$ [Eq. (\ref{G0})] for the conduction electron is expressed as follows:
\beqa
G^{(0)}_{\mib{k}} (i\epsilonn) &=&
\sum_{\gamma=\pm}\frac{A^\gamma_{\mib{k}}}{i\epsilonn-E^\gamma_{\mib{k}}},
\eeqa
where
\beqa
E_{\mib{k}}^{\gamma}&=&\frac{1}{2}\left[\ep_{\mib{k}}+E_0+\gamma\sqrt{\left(\ep_{\mib{k}}-E_0\right)^2+4aV^2/M}\right],
\eeqa
and
\beqa
A_{\mib{k}}^{\gamma}&=&\frac{E_{\mib{k}}^{\gamma}-E_0}{E_{\mib{k}}^{\gamma}-E_{\mib{k}}^{-\gamma}}.
\eeqa


Here, note that the self-energy of the conduction electrons of $O[ (1/N)^0]$, $\Sigma_{\mib{k}\tau\sigma}^{(0)}(\epsilon+i0_+)$, does not have an imaginary part at $\epsilon=0$ nor an incoherent part away from the Fermi level, so that only a quasiparticle band is formed and the system behaves as a Fermi liquid.
In order to obtain those terms arising from incoherent contributions, we need to take into account the contributions of $O( 1/N)$, as discussed in the subsections below.

\subsection{Vertices of $O[ (1/N)^0]$}

\begin{figure}
\includegraphics[width=10cm]{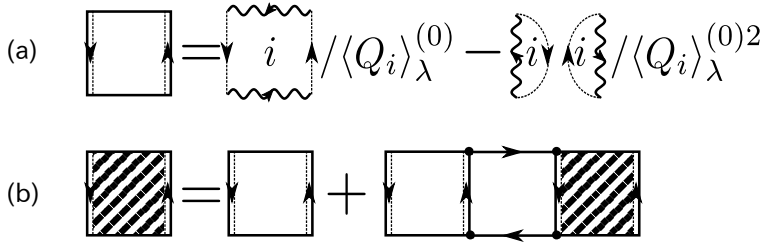}
\caption{(a) The local vertex $\Gamma^{{\rm loc}(0)}$ of $O[ (1/N)^0]$, and (b) the full vertex $\Gamma^{(0)}_{\mib{q}}$ of $O[ (1/N)^0]$.}
\label{Fig:local0}
\end{figure}

In order to calculate the self-energy of $O(1/N)$, we need the full vertices up to $O[ (1/N)^0]$.
The diagrams contributing to the local vertices of $O[ (1/N)^0]$ are shown in Fig. \ref{Fig:local0}(a), in which there is no summation with respect to the wave vectors.
$\Gamma^{{\rm loc}(0)}(i\epsilon_{n1}, i\epsilon_{n2}; i\nun)$ denotes the sum of these contributions, which is given by
\beqa
&&\Gamma^{{\rm loc}(0)}(i\epsilon_{n1}, i\epsilon_{n2}; i\nun)\non\\
&=&\Gamma^{{\rm loc}(0A)}(i\epsilon_{n1}, i\epsilon_{n2}; i\nun)+\Gamma^{{\rm loc}(0B)}(i\epsilon_{n1}, i\epsilon_{n2}; i\nun),\label{localv0}
\eeqa
where the vertex $\Gamma^{{\rm loc}(0A)}$ includes no singular term,
\beqa
&&\Gamma^{{\rm loc}(0A)}(i\epsilon_{n1}, i\epsilon_{n2}; i\nun)\non\\
&=&\frac{a^2V^4}{M}\left[\frac{\bar{F}(-i\nun)}{(i\epsilon_{n1}-E_0)(i\epsilon_{n2}-E_0)}\right.\non\\
&&\hspace{3em}\left.+\frac{\bar{F}(i\nun)}{(i\epsilon_{n1}+i\nun-E_0)(i\epsilon_{n2}+i\nun-E_0)}\right]\delta_{\sigma_1,\sigma_4}\delta_{\sigma_2,\sigma_3},\non\\
\eeqa
where
\beqa
\bar{F}(z)=\frac{1}{a}F^{(0)}_{i\sigma}(z+\lambda_i+\ep_{\Gamma_3}-E_0),\label{Eq:F}
\eeqa
while the vertex $\Gamma^{{\rm loc}(0B)}$ includes a singular term proportional to $T^{-1}$,
\beqa
\Gamma^{{\rm loc}(0B)}(i\epsilon_{n1}, i\epsilon_{n2}; i\nun)&=&-\frac{a^2V^4}{M}\frac{1}{T}\delta_{\nun,0}\frac{1}{(i\epsilon_{n1}-E_0)(i\epsilon_{n2}-E_0)}\non\\
&&\times\left(\delta_{\sigma_1,\sigma_4}\delta_{\sigma_2,\sigma_3}-M^{-1}\delta_{\sigma_1, \sigma_3}\delta_{\sigma_2, \sigma_4}\right).\label{gammalocal0b}
\eeqa
Note that $\Gamma^{{\rm loc}(0B)}$ vanishes in the single-channel case because $\sigma_1=\sigma_2=\sigma_3=\sigma_4$ and $M=1$.

The origin of a singularity proportional to $1/T$ in Eq. (\ref{gammalocal0b}) is understood as follows.
For $\nu_n=0$, the first diagram of the r.h.s. in Fig. \ref{Fig:local0} (a), which we denote as $\Gamma^{{\rm loc}(01)}$, is explicitly given as
\beqa
&&\Gamma^{{\rm loc}(01)}(i\epsilon_{n1}, i\epsilon_{n2}; 0)\non\\
&=&-V^4T\sum_{\epsilon_n}
F^{(0)}_{i\sigma_1}(i\epsilon_n)
F^{(0)}_{i\sigma_2}(i\epsilon_n)\non\\
&&\hspace{2cm}\times B^{(0)}_{i\tau_1}(i\epsilon_n+i\epsilon_{n1})
B^{(0)}_{i\tau_2}(i\epsilon_n+i\epsilon_{n2})
/\langle {\hat Q}_i\rangle_{\lambda}^{(0)}
\delta_{\sigma_1,\sigma_4}\delta_{\sigma_2,\sigma_3}\non\\
&=&-V^4\frac{-1}{2\pi i}\oint dz e^{-z/T}
\left[\frac{a}{z-\lambda_i-(\ep_{\Gamma_3}-E_0)}+C^{(0)}(z)\right]^2\non\\
&&\times\frac1{i\epsilon_n+i\epsilon_{n1}-\lambda_i-\ep_{\Gamma_3}}
\frac1{i\epsilon_n+i\epsilon_{n2}-\lambda_i-\ep_{\Gamma_3}}
M^{-1}e^{(\lambda_i+\ep_{\Gamma_3}-E_0)/T}
\delta_{\sigma_1,\sigma_4}\delta_{\sigma_2,\sigma_3}.\label{gammaloc01}
\eeqa
The singular factor $1/T$ arises from differentiating the term $e^{-z/T}$ with respect to $z$ in the process of the contour integration around the double pole $\{a/[z-\lambda_i-(\ep_{\Gamma_3}-E_0)]\}^2$ in the last line of Eq. (\ref{gammaloc01}).
This singular vertex corresponds to the ballistic scattering in the sense that the energy transfer $\nu_n$ is vanishing.

The full vertex of $O[ (1/N)^0]$, which contributes to the imaginary part of the self-energy of conduction electrons, is defined by the Bethe-Salpeter equation illustrated in Fig.~\ref{Fig:local0}(b),
because the number of the sum of the wave vector and the number of the sum of $\tau$ are the same in each diagram.

By denoting the full vertex of $O[ (1/N)^0]$ as $\Gamma^{(0)}_{\mib q}(i\epsilon_{n1}, i\epsilon_{n2}; i\nun)$, we obtain
\beqa
&&\hspace{-2em}\Gamma^{(0)}_{\mib q}(i\epsilon_{n1}, i\epsilon_{n2}; i\nun)\non\\
&&\hspace{-3em}=\Gamma^{(0A)}_{\mib q}(i\epsilon_{n1}, i\epsilon_{n2}; i\nun)+\Gamma^{(0B)}_{\mib q}(i\epsilon_{n1}, i\epsilon_{n2}; i\nun),
\eeqa
where $\Gamma^{(0A)}_{\mib q}$ makes a contribution in both cases of a single channel and a multichannel, and has no singular term,
\beqa
&&\Gamma^{(0A)}_{\mib q}(i\epsilon_{n1}, i\epsilon_{n2}; i\nun)\non\\
&=&\delta_{\sigma_1,\sigma_4}\delta_{\sigma_2,\sigma_3}\frac{a^2V^4}{M}\frac{1}{K_{\mib q}(i\nun)}\non\\
&&\times\left\{\frac{\bar{F}(i\nun)}{i\epsilon_{n1}+i\nun-E_0}\left[\frac{1+\bar{F}(-i\nun)f_{\mib q}^{(0,2)}(i\nun)}{i\epsilon_{n2}+i\nun-E_0}-\frac{\bar{F}(-i\nun)f_{\mib q}^{(1,1)}(i\nun)}{i\epsilon_{n2}-E_0}\right]\right.\non\\
&&\hspace{.5em}\left.-\frac{\bar{F}(-i\nun)}{i\epsilon_{n1}-E_0}\left[\frac{\bar{F}(i\nun)f_{\mib q}^{(1,1)}(i\nun)}{i\epsilon_{n2}+i\nun-E_0}-\frac{1+\bar{F}(i\nun)f_{\mib q}^{(2,0)}(i\nun)}{i\epsilon_{n2}-E_0}\right]\right\},
\eeqa
while the vertex $\Gamma^{(0B)}_{\mib q}$ makes a contribution only in the multichannel case and consists of a singular term:
\beqa
&&\Gamma^{(0B)}_{\mib q}(i\epsilon_{n1},i\epsilon_{n2};i\nun)\non\\
&=&-(\delta_{\sigma_1,\sigma_4}\delta_{\sigma_2,\sigma_3}-M^{-1}\delta_{\sigma_1,\sigma_3}\delta_{\sigma_2,\sigma_4})\frac{a^2V^4}{M}\frac{1}{T}\delta_{\nun,0}\non\\
&&\times\frac{1}{i\epsilon_{n1}-E_0}\left[1-f_{\mib q}^{(0,3)}(0)+\frac{f_{\mib q}^{(0,2)}(0)}{i\epsilon_{n1}-E_0}\right]
\frac{1}{i\epsilon_{n2}-E_0}\left[1-f_{\mib q}^{(0,3)}(0)+\frac{f_{\mib q}^{(0,2)}(0)}{i\epsilon_{n2}-E_0}\right]\non\\
&&\times\frac{1}{K_{\mib q}(0)\left[K_{\mib q}(0)-T^{-1}f_{\mib q}^{(0,2)}(0)\right]},\label{fullv0}
\eeqa
where
\beqa
&&f_{\mib q}^{(l, m)}(i\nun)=\frac{a^2V^4}{M}T\sum_{\epsilonn}\frac{1}{(i\epsilonn+i\nun-E_0)^l(i\epsilonn-E_0)^m}\non\\
&&\hspace{5em}\times\frac{1}{N_L}\sum_{\sigma,\mib{k}}G_{\mib{k}+\mib{q}}(i\epsilonn+i\nun)G_{\mib{k}}(i\epsilonn),\label{Eq:flm}\\
&&K_{\mib q}(i\nun)\non\\
&=&1+\bar{F}(-i\nun)f_{\mib q}^{(0,2)}(i\nun)+\bar{F}(i\nun)f_{\mib q}^{(2,0)}(i\nun)\non\\
&&+\bar{F}(-i\nun)\bar{F}(i\nun)\left\{f_{\mib q}^{(0,2)}(i\nun)f_{\mib q}^{(2,0)}(i\nun)-\left[f_{\mib q}^{(1,1)}(i\nun)\right]^2\right\}.\label{K}
\eeqa
Although the explicit expression of Eq. (\ref{K}) is rather lengthy,
$K_{\mib{q}}(0)$ is given by a relatively simple form as
\beqa
K_{\mib{q}}(0)&=&\left[1-f_{\mib q}^{(0,3)}(0)\right]^2
-f_{\mib{q}}^{(0,2)}(0)f_{\mib{q}}^{(0,4)}(0)\non\\
&&+\frac{2}{a}C(\lambda+\ep_{\Gamma_3}-\ep_{\Gamma_7}-E_0)f_{\mib q}^{(0,2)}(0).
\label{K0}
\eeqa
Here, note that the second term in the square bracket in the denominator in the last term of the r.h.s. of Eq. (\ref{fullv0}) includes a singular term that is proportional to $T^{-1}$.
Details of the calculations of the vertices $\Gamma^{\rm{loc}(0)}$ and $\Gamma_{\mib{q}}^{(0)}$ are given in Sect. 3 of Ref. \citen{Tsuruta5}.

\subsection{Self-energy of $O(1/N)$}

\begin{figure}
\includegraphics[width=10cm]{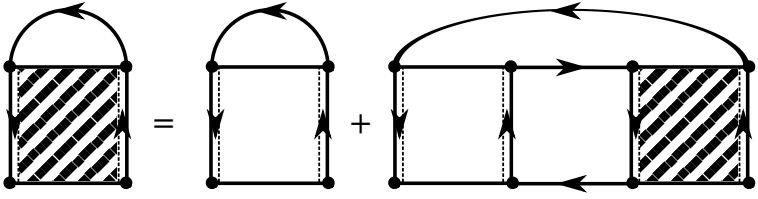}
\caption{
Feynman diagram representation of the self-energy $\Delta\Sigma_{\mib{k}\tau\sigma}^{(1)}(i\epsilonn)$.
This is based on that representing the Bethe-Salpeter equation for the full vertex $\Gamma$ shown in Fig. 3(b).
The striped square represents the full vertex $\Gamma$ of $O[ (1/N)^0]$, which consists of the local part depicted by the open square, and the part giving the intersite effect through the particle--hole propagation of conduction electrons.
}
\label{Fig:Sigma1}
\end{figure}

Now, we calculate the self-energy of conduction electrons up to $O(1/N)$ in the two-channel case ($M=2$) to prove that the imaginary part of the self-energy does not satisfy the Fermi liquid relation even though the system retains the translational symmetry.
Let us denote the contribution of the Feynman diagram shown by the l.h.s. of the diagrammatic relation in Fig. \ref{Fig:Sigma1} as $\Delta\Sigma^{(1)}_{\mib{k}\tau\sigma}$, which is on the order of $O(1/N)$ according to the rule [Eq. (\ref{O1N})], because there is one summation of the wave vector and no summation of $\tau$,
and the full vertex $\Gamma$ depicted by a striped square is of $O[ (1/N)^0]$ in the sense of the $1/N$-expansion formalism.
Its nonzero imaginary part is given in the following form:
\beqa
&&\hspace{1em}\Im\Delta\Sigma^{(1)}_{\mib{k}\tau\sigma}(\epsilon+i0_+)
=\Im\Delta\Sigma^{(1{\rm FL})}_{\mib{k}\tau\sigma}(\epsilon+i0_+)+\Im\Delta\Sigma^{(1{\rm NFL})}_{\mib{k}\tau\sigma}(\epsilon+i0_+),\label{ISigma1}
\eeqa
where ${\rm Im}\Delta\Sigma^{(1{\rm FL})}_{\mib{k}\tau\sigma}$ is the imaginary part of the self-energy of the Fermi liquid type:
\beqa
&&\Im\Delta\Sigma^{(1{\rm FL})}_{\mib{k}\tau\sigma}(\epsilon+i0_+)\non\\
&=&-\pi\frac{1}{N_L^2}\sum_{\sigma,\mib{k}_{1},\mib{k}_{2}}\sum_{\gamma_1,\gamma_2,\gamma_3}A_{\mib{k}_{1}}^{\gamma_1}A_{\mib{k}_{2}}^{\gamma_2}A_{\mib{k}-\mib{k}_{1}+\mib{k}_{2}}^{\gamma_3}\non\\
&&\times\left|\Gamma_{\mib{k}_{2}-\mib{k}_{1}}^{(0A)}(\epsilon,E_{\mib{k}_{1}}^{\gamma_1};E_{\mib{k}_{2}}^{\gamma_2}-E_{\mib{k}_{1}}^{\gamma_1})\right|^2\non\\
&&\times\delta\left(\epsilon-E_{\mib{k}_{1}}^{\gamma_1}+E_{\mib{k}_{2}}^{\gamma_2}-E_{\mib{k}-\mib{k}_{1}+\mib{k}_{2}}^{\gamma_3}\right)\non\\
&&\times\frac{\cosh\left(\epsilon/2T\right)}{4\cosh\left(E_{\mib{k}_{1}}^{\gamma_1}/2T\right)\cosh\left(E_{\mib{k}_{2}}^{\gamma_2}/2T\right)\cosh\left(E_{\mib{k}-\mib{k}_{1}+\mib{k}_{2}}^{\gamma_3}/2T\right)}\\
&\simeq&-\pi\alpha^{(1)}_{\mib{k}}\left[\epsilon^2+(\pi T)^2\right],\label{eq:sigmafl}
\eeqa
and ${\rm Im}\Delta\Sigma^{(1{\rm NFL})}_{\mib{k}\tau\sigma}$ is the imaginary part of the self-energy of the non-Fermi liquid type which does not appear in the single-channel case,
\beqa
&&\hspace{1em}\Im\Delta\Sigma^{(1{\rm NFL})}_{\mib{k}\tau\sigma}(\epsilon+i0_+)\non\\
&&=-\pi \left(\frac{aV^2}{\epsilon-E_0}\right)^2\frac{1}{N_L}\sum_{\mib{q}}\frac{\rho_{\mib{k}+\mib{q}}(\epsilon)}{K_{\mib{q}}(0)-T^{-1}f^{(0, 2)}_{\mib{q}}(0)}\non\\
&&\hspace{2em}\times\frac{1}{K_{\mib{q}}(0)}\left[1-f^{(0, 3)}_{\mib{q}}(0)+\frac{f^{(0,2)}_{\mib{q}}(0)}{\epsilon-E_0}\right]^2\left(1-\frac{1}{M^2}\right).\non\\
&&\label{ISigma1bM}
\eeqa
The coefficient $\alpha_{\mib{k}}^{(1)}$ in Eq. (\ref{eq:sigmafl}) is a constant on the order of $V^2N_{\rm F}/E_0^2$,
and $\rho_{\mib{k}}(\epsilon)$ in Eq. (\ref{ISigma1bM}) is the spectral function of conduction electrons,
\beqa
\rho_{\mib{k}\tau\sigma}(\epsilon)=-\frac{1}{\pi}{\rm Im}G_{\mib{k}\tau\sigma}(\epsilon+i0_+),\label{Eq:sw}
\eeqa
which depends on the wave vector $\mib{k}$ and thus on the band structure of conduction electrons, in general.
We also calculated the real part of the self-energies given by Fig. \ref{Fig:Sigma1}.
There are other diagrams of $O(1/N)$ shown in Figs. 6(a) and (c)-(k) in Ref. \citen{Tsuruta5}, which contribute to the self-energy, but they only modify $E_0$ and the chemical potential. Therefore, we neglect those diagrams.
As a result, in the region $|\epsilon|\ll E_0$ and $T\ll E_0$, the self-energy of $O(1/N)$ is given as
\beqa
\Sigma_{\mib{k}\tau\sigma}(\epsilon+i0_+)&=&\frac{1}{M}\frac{\tilde{a}_{\mib{k}}V^2}{\epsilon-\tilde{E}_{0\mib{k}}}-\Delta\mu_{\mib{k}}(\epsilon)\non\\
&&-\pi i\alpha^{(1)}_{\mib{k}}\left[\epsilon^2+\left(\pi T\right)^2\right]-\pi i\frac{\beta^{(1)}_{\mib{k}}}{1+T^*_{\mib{k}}/T}(1-M^{-2}),\label{Eq:sigmatotal}
\eeqa
where the imaginary part is the approximate expression with the high accuracy of that given by Eq. (\ref{ISigma1bM}), as verified numerically.
The coefficients $\alpha^{(1)}_{\mib{k}}$ and $\beta^{(1)}_{\mib{k}}$ are proportional to $N_{\rm F}V^2T_{\rm K}^{-2}$ and $N_{\rm F}^{-1}$, respectively.

\section{Chemical Potential}

\begin{figure}
\includegraphics[width=10cm]{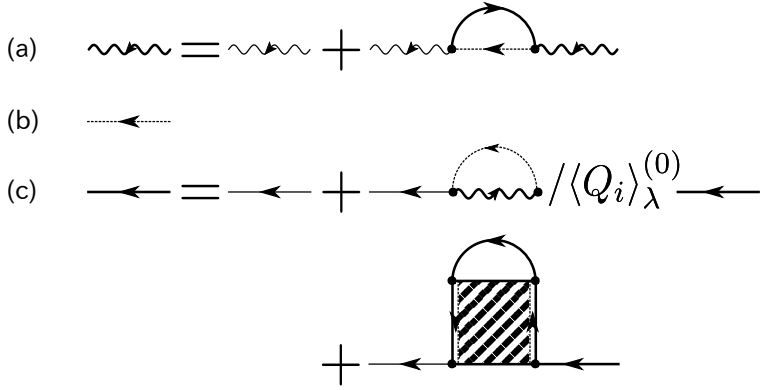}
\caption{Feynman diagram representation of the Dyson equations for the single-particle Green functions on the order of $O(1/N)$: (a) $F_{i\sigma}^{(1)}(i\epsilon_n)$ for the slave fermion, (b) $B_{i\tau}^{(1)}(i\nu_n)$ for the pseudoboson, and (c) $G_{\mib{k}\tau\sigma}^{(1)}(i\epsilon_n)$ for the conduction electron.}
\label{Fig:N1}
\end{figure}

In this section, we discuss how the $T$ dependence of chemical potential, $\mu(T)$, is calculated.
To this end, we first determine the Green functions up to the order of $O(1/N)$ by solving the Dyson equations shown diagrammatically in Fig. \ref{Fig:N1}.

$\mu(T)$ is determined by solving the following self-consistent equations:
\beqa
&&n_{\rm tot}=n_c+n_f,\label{sce1}\\
&&n_c=\frac1M\sum_{\sigma=1}^M\sum_{\tau=1}^N\frac{1}{N_L}\sum_{\mib{k}}\int_{-\infty}^{\infty}d\epsilon
\left[-\frac{1}{\pi}{\rm Im}G^{(1)}_{\mib{k}\tau\sigma}(\epsilon+i0_+)\right]
f(\epsilon),\\
&&n_f=(2n_{\rm pseudoboson}+n_{\rm slave\mathchar`-fermion})/2,\label{Eq:nf1}
\eeqa
where
$n_{\rm tot}$ is the total electron number per site and per channel, 
$n_{c}$ is the conduction electron number per site and per channel, 
$n_{\rm pseudoboson}$ is the number of pseudobosons per site,
$n_{\rm slave\mathchar`-fermion}$ is the number of slave fermions per site,
and $n_{f}$ is the $f$-electron number per site and per channel,
and is given by 
\beqa
n_f=1-\frac{a}{2}.\label{Eq:nf2}
\eeqa
In deriving Eq. (\ref{Eq:nf2}) from Eq. (\ref{Eq:nf1}), we have used the relations
$n_{\rm pseudoboson}=1-a$ and $n_{\rm slave\mathchar`-fermion}=a$.
Now, instead of Eqs. (\ref{Eq:E00}) and (\ref{Eq:a0}),
the residue $a$ and the binding energy $E_0$ of the slave fermion
should be determined up to the order of $O(1/N)$ by solving the coupled self-consistent equations
\beqa
 \ep_{\Gamma_3}-\ep_{\Gamma_7}-E_0 -\frac{V^2}{N_{\rm L}}\sum_{\mib{k},\tau}
 \int d\epsilon f(\epsilon)\frac{-1}{\pi}{\rm Im}G^{(1)}_{\mib{k}\tau\sigma}(\epsilon+i0_+)\frac{1}{\epsilon-E_0}=0,\label{Eq:E01}
\eeqa
and
\beqa
\frac1a= 1+\frac{V^2}{N_{\rm L}}\sum_{\mib{k},\tau}
 \int d\epsilon f(\epsilon)\frac{-1}{\pi}{\rm Im}G^{(1)}_{\mib{k}\tau\sigma}(\epsilon+i0_+)\frac{1}{(\epsilon-E_0)^2},
\eeqa
where the retarded Green function of conduction electrons is given by
\beqa
G^{(1)}_{ {\rm k}\tau\sigma}(\epsilon+i0_+)
=\frac1{\epsilon+i0_+-\xi_{ {\rm k}}-\Sigma_{ {\rm k}\tau\sigma}(\epsilon+i0_+)},\label{Eq:G1}
\eeqa
with the retarded self-energy given by
\beqa
\Sigma_{ {\rm k}\tau\sigma}(\epsilon+i0_+)
=
\Sigma^{(0)}_{ {\rm k}\tau\sigma}(\epsilon+i0_+)
+\Delta\Sigma^{(1)}_{ {\rm k}\tau\sigma}(\epsilon+i0_+),\label{sce2}
\eeqa
where 
$\Sigma^{(0)}_{ {\rm k}\tau\sigma}(\epsilon+i0_+)$ and 
$\Delta\Sigma^{(1)}_{ {\rm k}\tau\sigma}(\epsilon+i0_+)$ are self-energies given by Eqs. (\ref{Sigma0}) and (\ref{ISigma1})-(\ref{ISigma1bM}), respectively.
The Green function for the slave fermion on the order of $O(1/N)$, $F^{(1)}_{i\sigma}$, is given by
\beqa
F^{(1)}_{i\sigma}(i \epsilonn ) &=& \frac{a}{i\epsilonn -\lambda_i -\left(\ep_{\Gamma_3} -E_0\right)}+C^{(1)}(i\epsilonn), \label{F1}
\eeqa
where
\beqa
C^{(1)}(i\epsilonn)&=&a(i\epsilonn-\lambda-\ep_{\Gamma_3}+\ep_{\Gamma_7}+E_0)\frac{V^2}{N_L}\sum_{\mib{k},\sigma}F_{i\sigma}^{(1)}(i\epsilonn)\non\\
&&\times\int d\epsilon f(\epsilon)\frac{-1}{\pi}{\rm Im}G^{(1)}_{\mib{k}\tau\sigma}(\epsilon+i0_+)\frac{1}{(\epsilon-E_0)^2(i\epsilonn+\epsilon-\lambda-\ep_{\Gamma_3}+\ep_{\Gamma_7})}.\non\\
\label{C1}
\eeqa
The relation $a\simeq E_0/N_{\rm F}V^2$ obtained up to the order of $O[ (1/N)^0]$ is not seriously altered even up to the order of $O(1/N)$,
which we have verified by numerical calculations.

In the numerical calculations below, for simplicity, we use the constant density of states $N(\epsilon)$ for conduction electrons given by
\beqa
N(\epsilon)=
\left\{
\begin{array}{ll}
\frac1D & (|\epsilon|\le D)\\
0 & (\mbox{otherwise}),\\
\end{array}
\right.
\eeqa
%
with parameters chosen as the total electron number $n_{tot}=2.0$, $D=1$, $V=0.3$, $\ep^{(0)}_{\Gamma_3}-\ep^{(0)}_{\Gamma_7}=-0.5$, $M=2$, and $N=2$, which lead to $E_0=0.0607$, $a=0.437$, $\mu=-0.332$, and the effective mass of the quasi-particle $m^*/m=6.35$.
Hereafter, we use this set of parameters, unless otherwise stated.

\begin{figure}
\includegraphics[width=10cm]{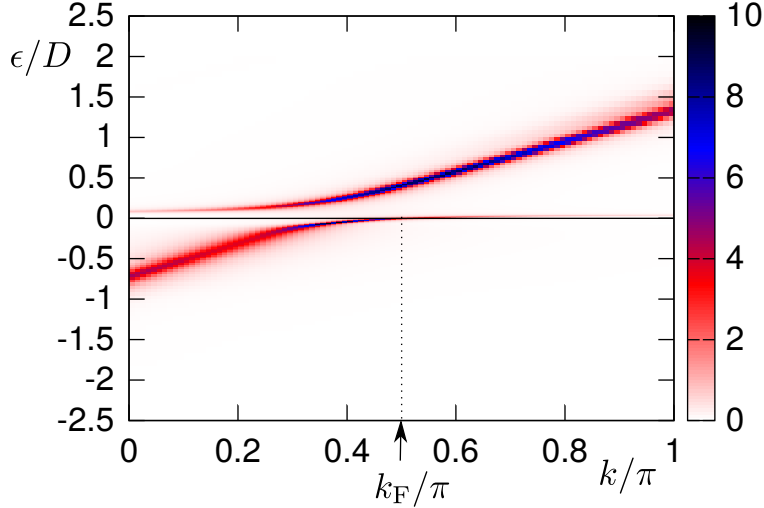}
\caption{Spectral weight $\rho_{\mib{k}\tau\sigma}(\epsilon)D=-{\rm Im}G^{(1)}_{\mib{k}\tau\sigma}(\epsilon+i0_+)D/\pi$ in the unit of $1/D$ at $T=0$ in $\epsilon$-$\mib{k}$ plane.
The Fermi wave number $k_{\rm F}$ is given by $k_{\rm F}/\pi=0.5$.
}
\label{Fig:spectralweight}
\end{figure}

Figure \ref{Fig:spectralweight} shows the spectral weight of conduction electrons, $\rho_{\mib{k}\tau\sigma}(\epsilon)$, defined by Eq. (\ref{Eq:sw}), at $T=0$.
At the Fermi level ($\epsilon=0$), the effective mass is heavy, and the peak is very sharp.
The localized state of conduction electrons at $\epsilon=\ep_{\Gamma_3}-\ep_{\Gamma_7}=-0.168D$ gives the spectral weight of conduction electrons a broad peak.

\begin{figure}
\includegraphics[width=8cm]{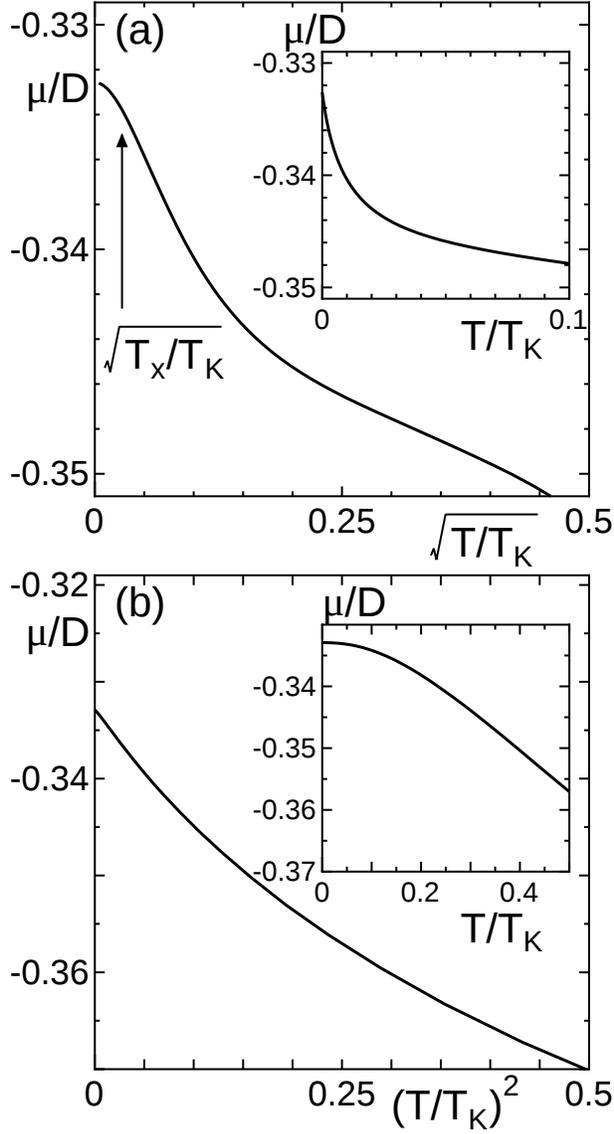}
\caption{$T$ dependence of chemical potential, $\mu(T)$, in the case of (a) two channels and (b) a single channel.
$T_x$ denotes the crossover temperature at which $\mu(T)$ changes from $\propto T$ to $\propto \sqrt{T}$ as $T$ increases.
}
\label{Fig:chem}
\end{figure}

By numerically solving self-consistent equations [Eqs. (\ref{sce1})-(\ref{sce2})], the chemical potential $\mu$ is determined as a function of $T$ and $n$.
Figure \ref{Fig:chem}(a) shows $\mu(T)$.
Note that $\mu(T)$ can be well fitted by the functional form as
\beqa
\mu=\mu_0\left(1-a_4\sqrt{T/E_0}\right),\label{cp1}
\eeqa
in the wide temperature region $T_x<T\ll T_{\rm K}$, where $T_x$ is a crossover temperature on the order of $0.0008T_{\rm K}$.
This should be compared with that in the case of a single channel, where the so-called Fermi liquid (FL) behavior is obtained for $\mu(T)$ at $T\ll T_{\rm K}$ without any crossover temperature scale other than $T_{\rm K}$:
\beqa
\mu=\mu_0\left[1-a_3(T/E_0)^2\right],
\eeqa
as shown in Fig. \ref{Fig:chem}(b).
The origin of the non-Fermi liquid (NFL) behavior in $\mu(T)$ [Eq. (\ref{cp1})]
can be traced back to the NFL $\epsilon$ and $T$ dependences of the self-energy [Eq. (\ref{ISigma1bM})].

Through this anomalous $T$ dependence of chemical potential, $\mu(T)$, various physical quantities turn out to have the anomalous NFL $T$ dependence, as shown in the following sections.


\section{Resistivity}
In this section, the $T$ dependence of electrical resistivity, $\rho(T)$, is discussed.
Although the conductivity (inverse of the resistivity) is, precisely speaking,
given by the current-current response function consisting of the fully renormalized Green function of conduction electrons and vertex corrections,\cite{YamadaYosida,Maebashi}
here, we adopt a simple approximation that the electrical resistivity $\rho$ is proportional to the imaginary part of the self-energy $\Sigma_{\mib{k}\tau\sigma}(i0_+;T)$ of conduction electrons at the Fermi wave vector,
because we are interested in the qualitative properties and general aspects of the two-channel Anderson lattice model.
Then, $\rho(T)$ is given as
%
\beqa
\rho&=&\rho_{\rm FL}+\rho_{\rm NFL},\label{Eq:rhototal}
\eeqa
where $\rho_{\rm FL}$ stems from the FL-type self-energy given by Eq. (\ref{eq:sigmafl}), and $\rho_{\rm NFL}$ stems from the NFL-type self-energy given by Eq. (\ref{ISigma1bM}).
Explicit expressions for $\rho_{\rm FL}$ and $\rho_{\rm NFL}$
are
\beqa
\rho_{\rm FL}&\simeq& r\ \pi N_{\rm F}V^2\frac{T^2}{E_0^2},\label{Eq:rhoFL}
\eeqa
and
\beqa
\rho_{\rm NFL}&\simeq& r\ \frac{a^2V^4}{E_0^2}\frac{1}{N_L}\sum_{\mib{q}}\frac{\rho_{\mib{k}+\mib{q}}(0)}{K_{\mib{q}}(0)-T^{-1}f^{(0, 2)}_{\mib{q}}(0)}\non\\
&&\hspace{2em}\times\left.\frac{1}{K_{\mib{q}}(0)}\left[1-f^{(0, 3)}_{\mib{q}}(0)-\frac{f^{(0,2)}_{\mib{q}}(0)}{E_0}\right]^2\left(1-\frac{1}{M^2}\right)\right|_{k=k_F},\label{eq:resistivity}
\eeqa
where $r$ is the ratio of the resistivity $\rho$ to $-{\rm Im}\Sigma_{k_F}(i0_+,T)$, which is the scattering rate of quasiparticles divided by the mass renormalization amplitude, and is reduced to $m/ne^2$ in the case of an isotropic Fermi liquid.

\begin{figure}
\includegraphics[width=10cm]{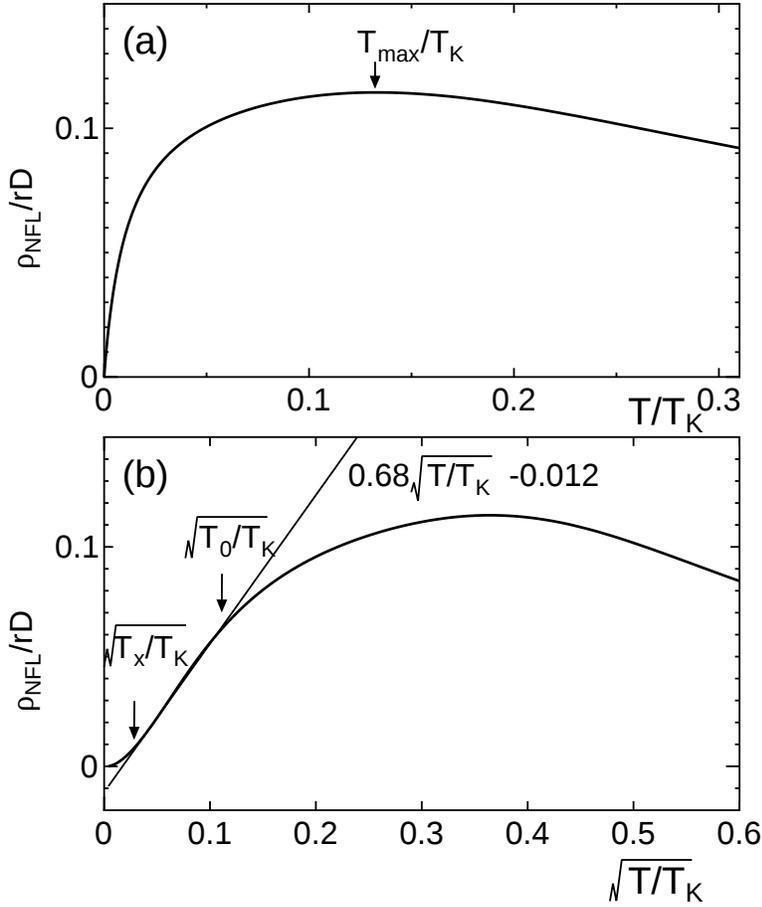}
\caption{
(a) $T/T_{\rm K}$ dependence of $\rho_{\rm NFL}/rD$, and
(b) $\sqrt{T/T_{\rm K}}$ dependence of $\rho_{\rm NFL}/rD$.
$T_{\rm max}$ denotes the temperature at which $\rho_{\rm NFL}/rD$
takes a maximum value.
$T_0$ denotes the crossover temperature at which the $T$ dependence of $\rho_{\rm NFL}/rD$
deviates from the $\sqrt{T}$ dependence to that with a much lower exponent as $T$ increases.
$T_x$ denotes the crossover temperature at which the $T$ dependence of $\rho_{\rm NFL}/rD$
changes from $\propto T$ to $\propto \sqrt{T}$ as $T$ increases.
}
\label{Fig:Im}
\end{figure}

\begin{figure}
\includegraphics[width=8cm]{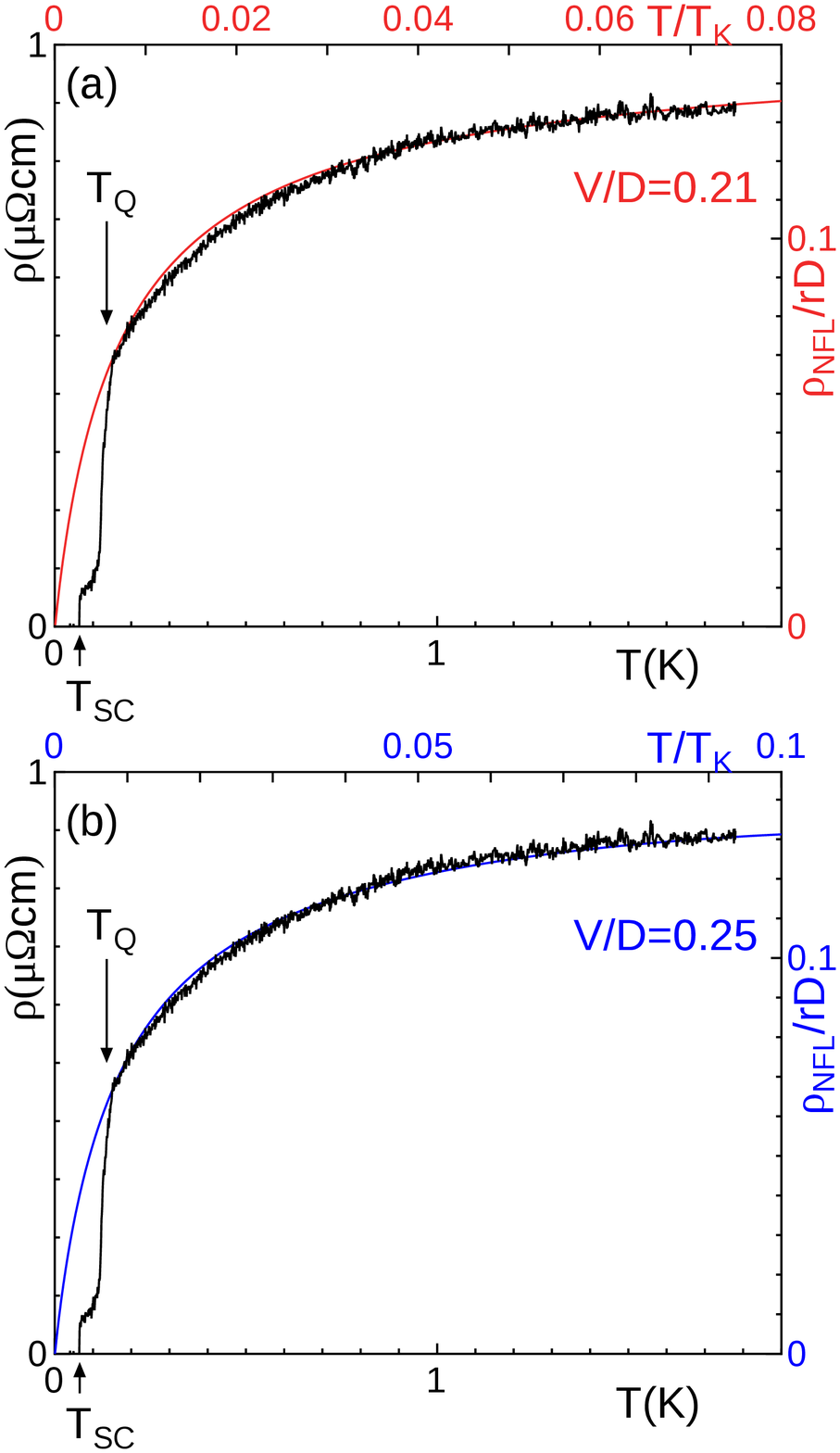}
\caption{
Comparison of $\rho(T)$ (black jagged curves) observed in PrIr$_2$Zn$_{20}$\cite{Onimaru} and theoretical results for $\rho_{\rm NFL}(T)$ in the case of (a) $V/D=0.21$ (red line) and (b) $V/D=0.25$ (blue line) given by Eq. (\ref{eq:resistivity}).
$T_Q$ and $T_{\rm SC}$ indicate the transition temperatures for the quadrupole ordering and superconducting ordering of PrIr$_2$Zn$_{20}$\cite{Onimaru}, respectively.
Note that the fundamental energy scale $T_{\rm K}$ is chosen as (a) $T_{\rm K}=24\ {\rm K}$ and (b) $T_{\rm K}=19\ {\rm K}$,
the ratio of the resistivity, $r$, is chosen as (a) $rD=6.67\ {\rm \mu\Omega cm}$ and (b) $rD=8.20\ {\rm \mu\Omega cm}$,
and the bandwidth $D$ is chosen as (a) $D=1220\ {\rm K}$ and (b) $D=540\ {\rm K}$.
}
\label{Fig:compare}
\end{figure}

Figure \ref{Fig:Im} shows the temperature dependence of 
$\rho_{\rm NFL}$ calculated from Eq. (\ref{eq:resistivity}) for conduction electrons for the same set of parameters mentioned above.
In the low-$T$ region, $T<T_x\simeq 0.00076T_{\rm K}$, the resistivity is proportional to T as seen from Eq. (\ref{eq:resistivity}),
while the $\sqrt{T}$ dependence can be seen in ${\rm Im}\Sigma_{k_F}(i0_+, T)$ in the region $0.00076\le T/T_{\rm K} \le 0.012$.
Recently, Sakai and Nakatsuji\cite{Sakai} and Onimaru et al.\cite{Onimaru} have experimentally found the $\sqrt{T}$ dependence of electrical resistivity in PrV$_2$Al$_{20}$ and PrIr$_2$Zn$_{20}$, respectively, in a rather wide temperature region above the quadrupolar ordering temperature $T_Q$.
We can fit our theoretical result to the experimental one in Ref. \citen{Onimaru}, as demonstrated in Fig. \ref{Fig:compare},
in which the theoretical results for $\rho_{\rm NFL}$ in the cases of (a) $V/D=0.21$ and (b) $V/D=0.25$ are shown.
The Kondo temperatures $T_{\rm K}$'s in these cases are $T_{\rm K}/D\simeq0.0197$ and $T_{\rm K}/D\simeq0.0351$, respectively, if other parameters are fixed as the same values adopted above.
In order to obtain the best fit in these cases, we have to adjust $D$, half the bandwidth of conduction electrons,
and the factor $r$, parameterizing the ratio of the resistivity and $-{\rm Im}\Sigma_{k_{\rm F}}(0,T)$; i.e., $D=1220\ {\rm K}$ ($T_{\rm K}=24\ {\rm K}$) and $rD=6.67\ \mu\Omega{\rm cm}$ in the case of (a), and
$D=540\ {\rm K}$ ($T_{\rm K}=19\ {\rm K}$) and $rD=8.20\ \mu\Omega{\rm cm}$ in the case of (b), respectively.

\begin{figure}
\includegraphics[width=10cm]{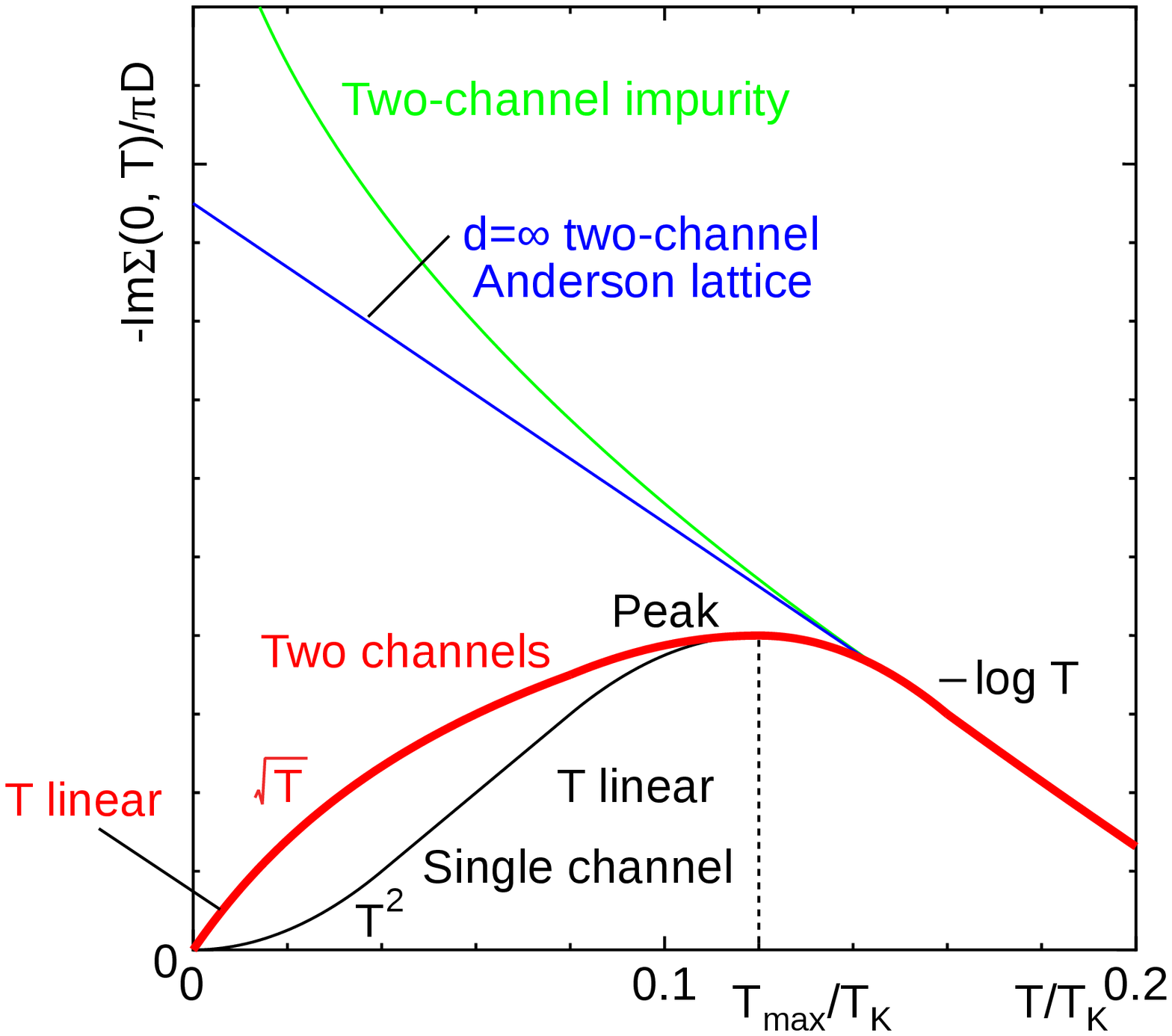}
\caption{Schematic $T$ dependence of $-{\rm Im}\Sigma(i0_+,T)$, in arbitrary unit,
in the case of single-channel Anderson lattice (black line),
two-channel impurity Anderson model (green line),\cite{Cox3}
two-channel Anderson lattice in $d\to\infty$ (blue line),\cite{Jarrell}
and two-channel Anderson lattice (red line).
Note that $d^2\rho/dT^2<0$ at 
$T<T_{\rm max}$ in the two-channel case, while $d^2\rho/dT^2>0$ at 
$T\ll T_{\rm max}$ in the single-channel case.
}
\label{Fig:STR}
\end{figure}

We can qualitatively understand this behavior from Fig. \ref{Fig:STR}.
The imaginary part of the self-energy of conduction electrons, $-{\rm Im}\Sigma(i0_+, T)$, is proportional to $-\log T$ in the high-temperature region $T\gtrsim T_{\rm K}$ and there appears a peak 
at $T=T_{\rm max}$, because the imaginary part has to be zero at $T=0{\rm K}$ in the lattice systems.
In the case of two channels, the second derivative is zero in the limit of $T\to 0$.
Therefore, the second derivative is negative between $T=0{\rm K}$ and $T=T_{\rm max}$.
This negative second derivative curve gives $-{\rm Im}\Sigma_{\mib{k}\tau\sigma}(\epsilon=0,T)$ an approximate $\sqrt{T}$ dependence at $T<T_{\rm K}$.
On the other hand, in the case of a single channel, it is proportional to $T^2$ in the limit of $T\to 0$.
Around the peak, the second derivative of $-{\rm Im}\Sigma_{\mib{k}\tau\sigma}(i0_+,T)$ is negative and it is positive near zero temperature.
Therefore, in the middle region of $T$, the quantity $-{\rm Im}\Sigma_{\mib{k}\tau\sigma}(i0_+,T)$ exhibits an approximate $T$-linear dependence.

\begin{figure}
\includegraphics[width=10cm]{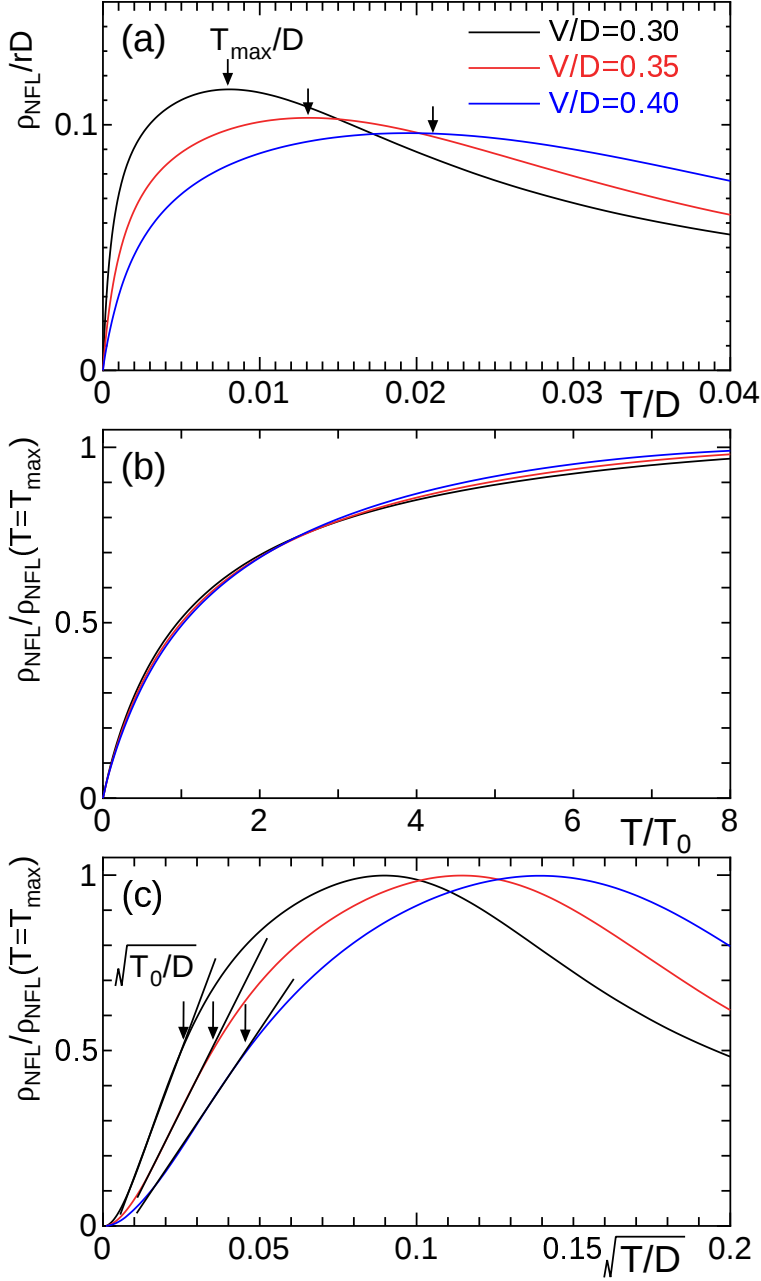}
\caption{(a) $T/D$ dependence of $\rho_{\rm NFL}$,
defining $T_{\rm max}$ at which $\rho_{\rm NFL}$takes a maximum value, as shown by the down arrow.
(b) $T/T_{0}$ dependence of $\rho_{\rm NFL}/\rho_{\rm NFL}(T=T_{\rm max})$
exhibiting an approximate scaling behavior.
(c) Definition of $T_0$ (shown by the down arrow) which is the crossover temperature from the $\sqrt{T}$ dependence to that with much lower exponent as $T$ increases.
}
\label{Fig:rhov}
\end{figure}

Figure \ref{Fig:rhov}(a) shows $\rho(T)$ for 
a series of 
hybridization $V$'s.
For a larger $V$, the slope in the limit $T\to0$ becomes smaller.
Figure \ref{Fig:rhov}(b) shows $T/T_0$ dependence of $\rho_{\rm NFL}$,
where $T_0$ is defined as the crossover temperature at which the $T$ dependence of $\rho_{\rm NFL}$ deviates from the $\sqrt{T}$ dependence to that with a much lower exponent as $T$ increases, as shown in Fig. \ref{Fig:rhov}(c).
This implies that $\rho(T)$'s for the systems with different $V$'s fit in with a single curve if $T$ is scaled by $T_0$, which is $V$-dependent, as shown in Fig. \ref{Fig:rhov}(c).
Namely, we obtain the scaling behavior $\rho(T)=R(T/T_0)$ with the single function $R(x)$ for different $T_0$'s. 
This scaling behavior in $\rho(T)$ has recently been observed in PrIr$_2$Zn$_{20}$\cite{Onimaru2} and PrV$_2$Al$_{20}$\cite{Nakatsuji}
under pressure, and also in PrIr$_2$Zn$_{20}$ under a magnetic field.\cite{Izawa}
As shown in Sect. 6, this type of scaling behavior also holds in the $T$ dependence of specific heat, $C(T)$, which is consistent with the experimental finding in PrIr$_2$Zn$_{20}$\cite{Onimaru2}.



According to Eqs. (\ref{Eq:rhototal})-(\ref{eq:resistivity}), we can rewrite the electrical resistivity at $T\lesssim T_{\rm max}\ll E_0$ as
\beqa
\rho=c_1\frac{T^2}{E_0^2}+\frac{c_2}{1+b\frac{E_0}{T}},\label{Eq:rho}
\eeqa
where the coefficients $c_1$ and $c_2$ are given by $c_1\simeq r\pi N_{\rm F}V^2$
and $c_2\simeq r(aV^2/E_0)^2N_{\rm F}\simeq r/N_{\rm F}$.
Here, note that $T_{\rm K}$ is the same quantity as $E_0$, as mentioned above.
The coefficient $b$ in the denominator of the second term of Eq. (\ref{Eq:rho}) is estimated as
$b\simeq 1/(1+N_{\rm F}E_0)$
because $K_{\mib q}(0)$ and $f_{\mib q}^{(0,2)}(0)$ in the denominator of Eq. (\ref{eq:resistivity}) are estimated as
$K_{\mib q}(0)\simeq 1+(1/N_{\rm F}E_0)$ and $f_{\mib q}^{(0,2)}(0)\simeq 1/N_{\rm F}$.
Since $c_1\simeq\pi(N_{\rm F}V)^2c_2$ and $N_{\rm F}V$ is moderately smaller than $1$ for heavy fermion systems, the NFL-type contribution, the second term of Eq. (\ref{Eq:rho}), dominates the FL-type one, the first term of Eq. (\ref{Eq:rho}), in the temperature region $T<T_{\rm max}\ll E_0$.

\begin{figure}
\includegraphics[width=10cm]{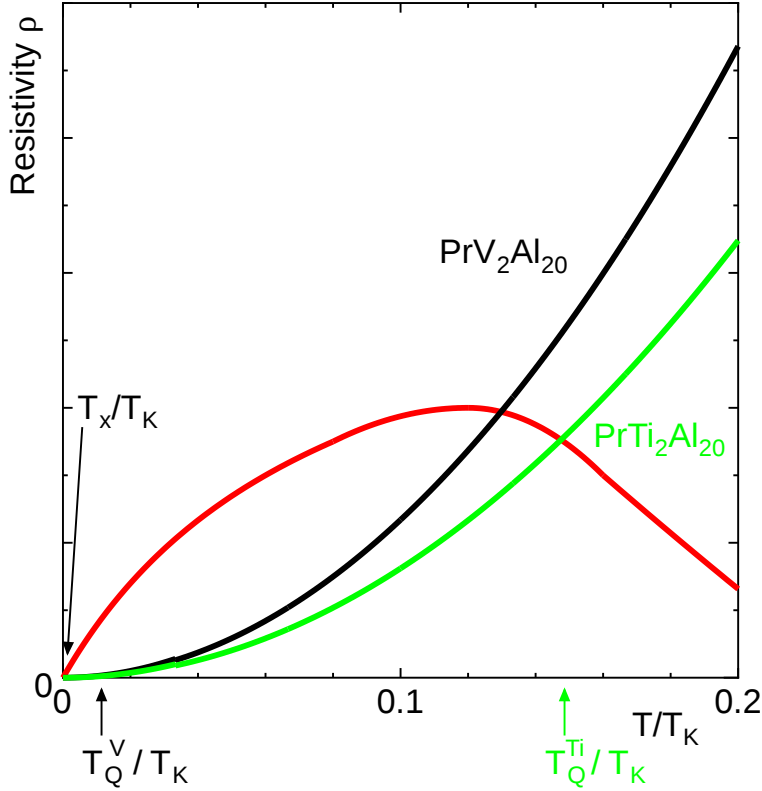}
\caption{Schematic $T$ dependences of NFL and FL contributions to the resistivity for PrV$_2$Al$_{20}$ and PrTi$_2$Al$_{20}$.
The red line indicates the NFL contribution shown in Fig. \ref{Fig:STR}, and the black and green lines indicate the FL contributions for PrV$_2$Al$_{20}$ and PrTi$_2$Al$_{20}$, respectively. 
$T_{Q}^{\rm V}$ and $T_{Q}^{\rm Ti}$ denote the quadrupole ordering temperatures of PrV$_2$Al$_{20}$ and PrTi$_2$Al$_{20}$, respectively.
}
\label{Fig:rhov2}
\end{figure}

Therefore, if $T_Q\ll T_{\rm max}$, the NFL-type $T$ dependence dominates the FL-type one at $T<T_{\rm max}$, as observed in PrV$_2$Al$_{20}$ and PrIr$_2$Zn$_{20}$.
On the other hand, if $T_Q\sim T_{\rm max}$, the FL-type $T$ dependence dominates the NFL-type one at $T>T_{\rm max}\sim T_Q$, as observed in PrTi$_2$Al$_{20}$.
Since the hybridization $V$ of PrTi$_2$Al$_{20}$ is smaller than that of PrV$_2$Al$_{20}$, the sister compound of PrV$_2$Al$_{20}$\cite{Sakai}, the $T_Q$ of PrTi$_2$Al$_{20}$ is about three times higher than that of PrV$_2$Al$_{20}$, i.e., $T_Q^{\rm Ti}\simeq2{\rm K}$ and $T_Q^{\rm V}\simeq0.6{\rm K}$, and the $E_0$ of PrTi$_2$Al$_{20}$ is expected to be much smaller than that of PrV$_2$Al$_{20}$ because $E_0$ has the sharp dependence on $V$ as $E_0\propto \exp({-{|\ep_{\Gamma_3}-\ep_{\Gamma_7}|}/{V^2N_F}})$.
Therefore, ($T_Q^{\rm Ti}/E_0^{\rm Ti}$) is expected to be much larger than ($T_Q^{\rm V}/E_0^{\rm V}$).
This situation is shown in Fig. \ref{Fig:rhov2}, explaining
the fact that the FL-type $\rho(T)$ is observed at $T>T_Q$ in PrTi$_2$Al$_{20}$, 
the sister compound of PrV$_2$Al$_{20}$
that exhibits the NFL-type $\rho(T)$ at $T>T_Q$,
as discussed above.

It should be noted that a different scenario explaining the difference of temperature dependence between PrV$_2$Al$_{20}$ and PrTi$_2$Al$_{20}$ was recently proposed from the viewpoint on the basis of competition between quadrupole and magnetic Kondo effect.\cite{KusunoseOnimaru}

\section{Difficulty in using DMFT for Multichannel Anderson Lattice Model}

In this section, the difficulty in using the DMFT when applied to calculating the transport properties of the two-channel Anderson or Kondo lattice model is discussed.

The DMFT is known to give a correct $T$ dependence of the resistivity, $\rho(T)$, in the single-channel Anderson lattice model: $\rho(T)\propto T^2$ at $T\ll T_{\rm K}$.
The DMFT explains the existence of the Mott transition in the Hubbard model,\cite{Georges}
and the Kondo insulator state and the heavy electron state in the Anderson lattice model.\cite{Jarrell3,Shimizu}
On the other hand, it gives an unphysical result in the two-channel Anderson or Kondo lattice model:
$\rho(T)\propto {\rm const.}$ in the limit $T\to0$.
Indeed, when the DMFT is applied to the two-channel Kondo lattice model with solvers of the impurity problem such as the quantum Monte Carlo method\cite{Jarrell} and continuous-time quantum Monte Carlo method\cite{Hoshino}, the resistivity is shown to be finite even at $T=0$.
However, in lattice systems that retain translational symmetry, the resistivity must be zero at $T=0$
owing to the conservation law of lattice momentum.\cite{YamadaYosida,Maebashi}
Such an insufficiency stems from the fact that the DMFT fails to properly take into account intersite effects.

This situation can be explicitly analyzed on the formalism of this paper.
The results suggest that the DMFT is insufficient for including the intersite effect.
Indeed, if we take the limit of the infinite spatial dimension, i.e., $d\to\infty$, of terms giving for the self-energy $\Delta\Sigma^{\rm (1NFL)}_{\mib{k}\tau\sigma}(\epsilon)$ [Eq. (\ref{ISigma1bM})], 
the second term on the r.h.s. of Fig. 4 (which manifests the intersite effects through the particle--hole propagation of the conduction electrons)
vanishes.
This is because such a term including the $n$ local vertices $\Gamma_{\rm loc}$'s, indicated by the open square, is of $O[d^{-(n-1)}]$.
Indeed, the factor $V^4$ appears through the hybridization between the conduction electrons and the local vertex $\Gamma_{\rm loc}$, and the wave vector summation of the particle--hole pair gives the factor $d$.
Then, considering $V=O(1/\sqrt{d})$ in the limit $d\to\infty$, the term in question is estimated as
\beqa
(V^4\Gamma_{\rm loc}d)^{n-1}\Gamma_{\rm loc}\sim (d^{-2}\cdot d)^{n-1}(\Gamma_{\rm loc})^n=d^{-(n-1)}(\Gamma_{\rm loc})^n.
\eeqa
Therefore, the contribution of the second term on the r.h.s. of Fig. 4 can be neglected in the limit $d\to\infty$.
On the other hand, the first term on the r.h.s. of Fig. 4 (which includes only the local vertex $\Gamma_{\rm loc}$ and manifests the local and on-site effects) remains nonvanishing even in the limit $d\to\infty$.

More explicitly, each quantity included in the r.h.s. of Eq. (\ref{ISigma1bM}) is estimated as in the following arguments.
According to the definition, Eq. (\ref{ISigma1bM}), the property $V=O(1/d)$,
and the fact that $(1/N_L)\sum_{\mib{k}}$ gives a factor proportional to the space dimension $d$, $f^{(l,m)}_{\mib{q}}(0)$'s given by Eq. (\ref{Eq:flm}) are estimated as
\beqa
f_{\mib{q}}^{(0,2)}(0)=f_{\mib{q}}^{(2,0)}(0)\sim\frac{a^2V^4}{E_0^2}\frac{N_{\rm F}d}{a}\frac{a^2}{a}\propto \frac{d^{\ -1}}{N_{\rm F}}\label{f02},\\
f_{\mib{q}}^{(0,3)}(0)\sim\frac{a^2V^4}{E_0^3}\frac{N_{\rm F}d}{a}\frac{a^2}{a}\propto \frac{d^{\ -1}}{E_0N_{\rm F}}\label{f03},
\eeqa
and
\beqa
f_{\mib{q}}^{(1,1)}(0)\sim\frac{a^2V^4}{E_0^2}\frac{N_{\rm F}d}{a}\frac{a}{a}\propto \frac{{d^{\ -1}}}{N_{\rm F}}\label{f11},
\eeqa
where we have used the relation $aV^2N_{\rm F}d\sim E_0$.
According to Eq. (\ref{C0}),
the $d$ dependence of $C(\lambda+\ep_{\Gamma_3}-\ep_{\Gamma_7}-E_0)$ is estimated as
\beqa
C(\lambda+\ep_{\Gamma_3}-\ep_{\Gamma_7}-E_0) \sim aV^2dN_{\rm F}\frac{a}{E_0^2} \propto \frac{d^{0}a}{E_0}.\label{F0d}
\eeqa
Then, with the use of Eqs. (\ref{f02})-(\ref{F0d}),
$K_{\mib{q}}(0)$ defined by Eq. (\ref{K0}) is estimated in the limit $d\to\infty$ as
\beqa
K_{\mib{q}}(0)= 1+O(d^{\ -1}).
\eeqa
Therefore, in the limit $d\to\infty$,
${\rm Im}\Delta\Sigma_{\mib{k}\tau\sigma}^{\rm (1NFL)}(\epsilon+i0_+)$
[Eq. (\ref{ISigma1bM})]
is given as
\beqa
&&{\rm Im}\Delta\Sigma^{\rm (1NFL)}_{\mib{k}\tau\sigma}(\epsilon)=-\pi\left(\frac{aV^2}{\epsilon-E_0}\right)^2\frac{1}{N_L}\sum_{\mib{k}}\rho_{\mib{k}\tau\sigma}(\epsilon)\left(1-M^{-2}\right).
\label{IMSigdinf}
\eeqa
A crucial point of the estimation above is that, in the limit $d\to\infty$, $T^{-1}f_{\mib{q}}^{(0,2)}(0)$
is neglected compared with $K_{\mib{q}}(0)$ in the denominator in the first line on the r.h.s. of Eq. (\ref{ISigma1bM}).

It is apparent that ${\rm Im}\Delta\Sigma_{\mib{k}\tau\sigma}^{(1)}$ given by Eq. (\ref{IMSigdinf}) remains finite even in the limit $T\to0$.
The above analysis, on the basis of the $1/N$-expansion approach, explicitly unveils the difficulty in explaining the results for the $\rho(T)$ obtained using the DMFT.
This shows that the intersite effect is important to obtain correct results in the two-channel Anderson or Kondo lattice model, and that the DMFT is not valid at least for discussing transport properties in the two-channel Anderson or Kondo lattice model.
On the other hand, in the case of the single-channel $(M=1)$ model,
${\rm Im}\Sigma^{\rm (1NFL)}_{\mib{k}\tau\sigma}(\epsilon)$ given by Eq. (\ref{ISigma1bM}) identically vanishes irrespective of $T$ so that the unphysical result does not appear, giving rise to a lack of difficulty.
The relationship among self-energies in the impurity model, $d=\infty$ lattice model (for which DMFT is valid), and the lattice model in finite dimensions (treated by the $1/N$-expansion formalism) is discussed in Appendix A.


\section{Specific Heat}
Under conventional experimental conditions, specific heat is measured with the fixed pressure $P$, not with the fixed volume $V$.
On the other hand, it is the specific heat with the fixed chemical potential $\mu$ and $V$ that is easily calculated on the basis of the conventional many-body theory based on the field theoretical method with Feynman diagrams.
Among these two specific heats, $C_P$ (with fixed $P$) and $C_V$ (with fixed $V$), the following relation holds:
\beqa
C_P=T\left(\frac{\partial S}{\partial T}\right)_P
=T\left(\frac{\partial S}{\partial T}\right)_V
\frac{\left(\frac{\partial P}{\partial V}\right)_S}
{\left(\frac{\partial P}{\partial V}\right)_T}
=C_V
\frac{\left(\frac{\partial P}{\partial V}\right)_S}
{\left(\frac{\partial P}{\partial V}\right)_T},
\label{sh0}
\eeqa
where $C_V$ is given by Eq. (\ref{cc}) in Appendix B.

With the use of the relations
of thermodynamic derivatives,
the derivative $(\partial P/\partial V)_S$ in Eq. (65) is transformed as
\beqa
\left(\frac{\partial P}{\partial V}\right)_S
&=&\frac{\partial (P,S)}{\partial (V,S)}
=\frac{\partial (P,S)}{\partial (V,T)}
\frac{\partial (V,T)}{\partial (V,S)}\non\\
&=&\left[\left(\frac{\partial P}{\partial V}\right)_T
\left(\frac{\partial S}{\partial T}\right)_V
-\left(\frac{\partial P}{\partial T}\right)_V
\left(\frac{\partial S}{\partial V}\right)_T
\right]
\frac1{\left(\frac{\partial S}{\partial T}\right)_V}\non\\
&=&\left(\frac{\partial P}{\partial V}\right)_T
\left[1-\frac{\left(\frac{\partial P}{\partial T}\right)_V}
{\left(\frac{\partial P}{\partial V}\right)_T}
\frac{
\left(\frac{\partial S}{\partial V}\right)_T}
{\left(\frac{\partial S}{\partial T}\right)_V}\right]\non\\
&=&\left(\frac{\partial P}{\partial V}\right)_T
\left[1+\left(\frac{\partial V}{\partial T}\right)_P\frac{T}{C_V}
\left(\frac{\partial S}{\partial V}\right)_T\right].
\eeqa
Therefore, the ratio $(\partial P/\partial V)_S/(\partial P/\partial V)_T$
in Eq. (\ref{sh0}) is given as
\beqa
\frac{\left(\frac{\partial P}{\partial V}\right)_S}
{\left(\frac{\partial P}{\partial V}\right)_T}
&=&1+T\left(\frac{\partial V}{\partial T}\right)_P\frac1{C_V}
\left(\frac{\partial S}{\partial V}\right)_T.\non\\
\label{sh2}
\eeqa

\begin{figure}
\includegraphics[width=10cm]{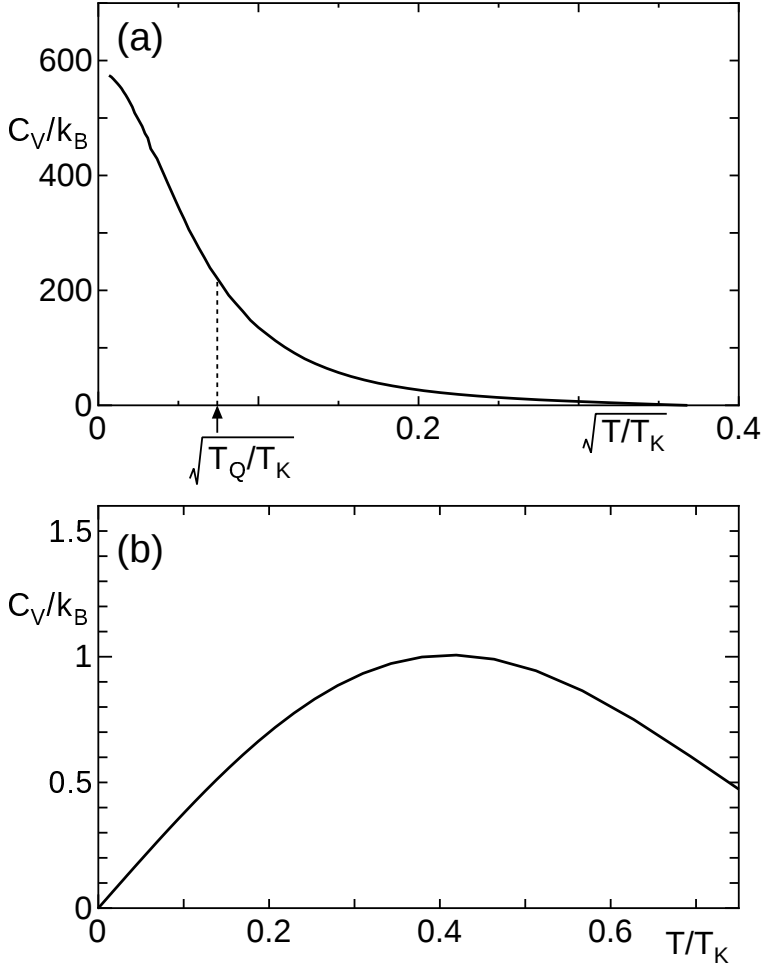}
\caption{$T$ dependence of specific heat, $C_V(T)$, for (a) two-channel case and (b) single-channel case.
$T_Q$ in (a) is the transition temperature for the antiferro-quadrupole ordering.
Note that the regions of $T/T_{\rm K}$ presented in (a) and (b) are quite different, so that $T/T_{\rm K}=0.1$ corresponds to $\sqrt{T/T_{\rm K}}\simeq0.3$.
}
\label{Fig:sh}
\end{figure}

The specific heat $C_V(T)$ is calculated using Eq. (\ref{cc}),
which exhibits the following $T$ dependence at $T_Q<T\ll E_0$, with $T_Q$ being the transition temperature of the quadrupole ordering:
\beqa
C_V(T)\simeq C_0\left(1-a_5\sqrt{T/E_0}\right),
\eeqa
as shown in Fig. \ref{Fig:sh}(a).
Therefore, by integrating the relation
$(\partial S/\partial T)_{V}=C_{V}(T)/T$
with respect to $T$, the $T$ dependence of entropy, $S(T)$, is given by
\beqa
S(T)\simeq S(T_Q)+C_0\left[\ln\frac{T}{T_Q}
-2a_5\left(\sqrt{\frac{T}{E_0}}-\sqrt{\frac{T_Q}{E_0}}\right)\right].
\eeqa
Then, the second term of Eq. (\ref{sh2}) is on the order of $O(T/E_0)$
within a logarithmic accuracy in $T$, because the fundamental energy scale determining the low-energy physics is given by $E_0$.
Therefore, considering that the $T$ dependence of physical quantities is scaled by $E_0$ at $T\ll E_0$, the second term of Eq. (\ref{sh2}) is neglected compared with the first term at $T\ll E_0$, giving $(\partial P/\partial V)_S/(\partial P/\partial V)_T\simeq1$.
Namely, $C_P(T)$ can be approximated by $C_V(T)$ as long as the $T$ dependence of the leading order, i.e., $O(\sqrt{T/E_0})$, is considered.

Figure \ref{Fig:sh} shows the $T$ dependence of specific heat, $C_V(T)$, for the two channels [(a)] and the single channel [(b)] in the unit of $k_{\rm B}$, the Boltzmann constant.
In the case of two channels ($M=2$), the specific heat $C_{V}(T)$ at $T_Q<T\ll E_0$ can be well fitted by
\begin{eqnarray}
C_V=C_0\left(1-a_5\sqrt{T/E_0}\right),\label{cv1}
\end{eqnarray}
as shown in Fig. \ref{Fig:sh}(a).
This $C(T)$ was observed in both PrV$_2$Al$_{20}$\cite{Sakai} and PrIr$_2$Zn$_{20}$\cite{Onimaru}.
$T_Q$ in Fig. \ref{Fig:sh}(a) is the transition temperature for the antiferro-quadrupole ordering and is estimated by RPA for the response function of pseudoboson density, which represents the quadrupole density arising from the $\Gamma_3$ non-Kramers doublet.
(Details of the calculation will be discussed elsewhere.)
However,
this result apparently breaks the third law of thermodynamics if it is applied down to $T=0$.
This deficiency may be avoided by taking into account the existence of ordered states such as quadrupole order, superconducting order, and magnetic order, while the quadrupole order is the most promising in the present model.
On the other hand, in the case of a single channel ($M=1$), the specific heat $C_V(T)$ is proportional to $T$, as shown in Fig. \ref{Fig:sh}(b), manifesting the Fermi liquid state.

\section{Magnetic Susceptibility of Conduction Electrons}
\begin{figure}
\includegraphics[width=10cm]{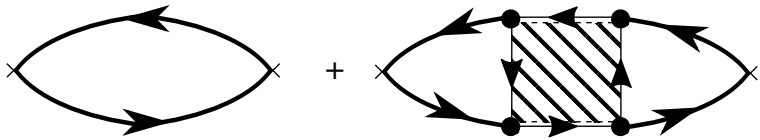}
\caption{Feynman diagrams giving the magnetic susceptibility.
The first and second terms represent $\chi_m^{(1)}$ and $\chi_m^{(2)}$, respectively.}
\label{Fig:diagramsus}
\end{figure}

In this section, we calculate the magnetic susceptibility $\chi_m$, which corresponds to the channel susceptibility 
in the present model given by Eqs. (\ref{Eq:Horg})-(\ref{Eq:Horg3}) or (\ref{H}) and (\ref{Q}).
The conduction electron component of the magnetic susceptibility described by the diagram shown in Fig. \ref{Fig:diagramsus} is larger than that from the slave fermion.
To the leading order in $1/N$, the $\chi_m$ is given explicitly by

\beqa
\chi_{m}(\mib{q})&=&\chi_m^{(1)}(\mib{q})+\chi_m^{(2)}(\mib{q}),\\
\chi_m^{(1)}(\mib{q})&=&-\mu_{\rm eff}^2 T\sum_{\epsilon_n}\frac{1}{N_L}\sum_{\mib{k},\tau}
G^{(1)}_{\mib{k}\tau\sigma}(i\epsilon_n)G^{(1)}_{\mib{k}+\mib{q}\tau\sigma}(i\epsilon_n),\\
\chi_m^{(2)}(\mib{q})
&=&-\mu_{\rm eff}^2 T^2\sum_{\epsilon_{n_1}, \epsilon_{n_2}}\frac{1}{N_L^2}\sum_{\mib{k}_1, \mib{k}_2, \tau}
G^{(1)}_{\mib{k}_1\tau\sigma}(i\epsilon_{n_1})G^{(1)}_{\mib{k}+\mib{q}\tau\sigma}(i\epsilon_{n_1})
\Gamma_{\mib{q}}^{(0)}(i\epsilon_{n_1}, i\epsilon_{n_2}; 0)\non\\
&&\times G^{(1)}_{\mib{k}_2\tau\sigma}(i\epsilon_{n_2})G^{(1)}_{\mib{k}+\mib{q}\tau\sigma}(i\epsilon_{n_2}),
\eeqa
where $\mu_{\rm eff}$ is the effective magnetic moment of the conduction electrons with $\Gamma_8$ symmetry.

\begin{figure}
\includegraphics[width=10cm]{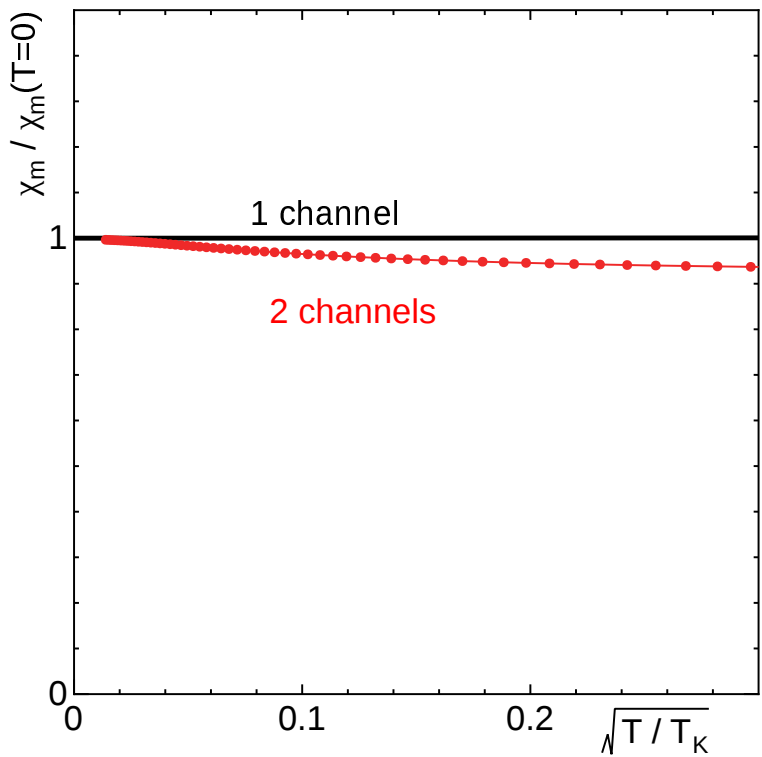}
\caption{$T$ dependence of magnetic susceptibility for two-channel case (red line with dots) and single-channel case (black line without dots).}
\label{Fig:sus}
\end{figure}

Figure \ref{Fig:sus} shows the $T$ dependence of $\chi_m=\chi_m(\mib{q}=0)$.
In the case of a single channel, $\chi_m$ is given in the form
\beqa
\chi_m=\chi_m(T=0)\left[1-a_6\left(T/E_0\right)^2\right].
\eeqa
In the case of the two channels, $\chi_m$ is given in the form
\beqa
\chi_m=\chi_m(T=0)\left(1-a_7\sqrt{T/E_0}\right).\label{Eq:chi}
\eeqa
In the non-Fermi liquid system,
the magnetic susceptibility is suppressed by the non-Fermi liquid imaginary part of the self-energy (ISE).
However, both the Fermi liquid ISE and the non-Fermi liquid ISE are zero at zero temperature.
Therefore, $\chi_m$ increases more rapidly in the non-Fermi liquid system than in the Fermi liquid system.
The theoretical result, Eq. (\ref{Eq:chi}), qualitatively explains the $T$ dependence in
PrV$_2$Al$_{20}$\cite{Sakai} and PrIr$_2$Zn$_{20}$\cite{Onimaru}. 
However, the cusp in the case of two channels at a low temperature is much smaller than that observed in experiments.

This quantitative discrepancy with the results of the experiments may be cured by taking into account the following two aspects that are not included in the model Hamiltonian Eq. (\ref{H}).
First, in our calculation, we have neglected the excited states in CEF of $4f^2$-configuration, including the magnetic degrees of freedom.
In other words, we have calculated the magnetic susceptibility of quasiparticles near the Fermi level.
However, there also arises the magnetic susceptibility through the virtual hopping processes including the excited CEF states,
i.e., 
the Van Vleck term, which gives a considerable contribution.
Indeed, it has been shown by numerical renormalization group (NRG) calculations~\cite{Yotsuhashi}
that the magnetic susceptibility of the impurity model for UPt$_3$ with the singlet CEF ground state in $f^2$-configuration consists of the Van Vleck-type contribution and that of quasiparticles,
the latter of which is not enhanced while the specific heat coefficient is highly enhanced.
It has also been shown by the slave-boson mean-field calculation~\cite{Ikeda}
that the quasiparticles' contribution to magnetic susceptibility is not enhanced if the CEF ground state is singlet in $f^2$-configuration even though the effective mass of quasiparticles is highly enhanced.

Secondly, it has been shown by the NRG calculation for the two-channel impurity Kondo model~\cite{Kusunose} that the perturbation breaking a particle--hole symmetry, such as the repulsive interaction among conduction electrons,
gives rise to the same divergent behavior in the channel susceptibility (the spin-orbital susceptibility in the present model) as the spin susceptibility (the quadrupolar susceptibility in the present model).
This aspect is expected to succeed in the case of the two-channel lattice system.


\section{Conclusions}

We have shown that the non-Fermi-liquid properties observed in PrV$_2$Al$_{20}$ and PrIr$_2$Zn$_{20}$ can be understood
on the basis of the two-channel Anderson lattice model with the use of the $1/N$-expansion formalism $\acute{\rm a}$ la Nagoya, which had been confirmed to be valid by properly taking strong correlation effects into account.
This Anderson lattice model simulates the heavy fermion systems with the $\Gamma_3$ non-Kramers doublet CEF ground state in $f^2$-configuration, such as
Pr$A_2$Al$_{20}$ ($A=$Ti, V) and PrIr$_2$Zn$_{20}$.
Results obtained in the present study are summarized as follows:\\
\\
1) The imaginary part of the conduction electrons, which has already been derived in Ref. \citen{Tsuruta4} as given by Eq. (1) up to the order of $O(1/N)$,
explains the non-Fermi-liquid $T$ dependence of resistivity, $\rho(T)$, observed in PrV$_2$Al$_{20}$ and PrIr$_2$Zn$_{20}$.
In particular, it turned out for the first time that $\rho(T)$ has a scaling form, $\rho(T)=R(T/T_{0})$, where $T_{0}$, the crossover temperature at which $\rho(T)$ changes from $\propto \sqrt{T}$ to $\propto T^\eta$ (with $\eta<1/2)$, depends on pressure through the change in the hybridization between $f$- and conduction electrons.\\
\\
2) It has been shown explicitly that the DMFT cannot correctly predict the behavior of $\rho(T)$ when applied to the multichannel Anderson lattice model
or multichannel Kondo lattice model 
because it is only valid in the limit of the infinite spatial dimension, $d\to\infty$, in which $T_{\mib{k}}^*$ in Eq. (1) vanishes in proportion to $1/d$, giving the unphysical result $\lim_{T\to 0}\rho(T)\ne 0$.
On the other hand, it works correctly when applied in the case of the single-channel Anderson lattice model with $M=1$, for which the second term in Eq. (1) vanishes, accidentally killing the unphysical effect.\\
\\
3) The $T$ dependence of chemical potential, $\mu(T)$, exhibits the $\sqrt{T}$ dependence as given by Eq. (\ref{cp1}) in a rather wide temperature region $T_x<T<0.02T_{\rm K}$, with the lower crossover temperature $T_x/T_{\rm K}\simeq 0.0008$
at which the $T$ dependence of $\rho(T)$ changes from $\propto T$ to $\propto \sqrt{T}$ as $T$ increases for the typical parameter set adopted, i.e., $V/D=0.3$.
The specific heat $C(T)$ and the magnetic susceptibility $\chi_m(T)$ exhibit similar $T$ dependences given by Eqs. (\ref{cv1}) and (\ref{Eq:chi}), respectively, or as shown in Table I.
The $T$ dependence of $C(T)$ is well fitted to the experimental results observed in PrV$_2$Al$_{20}$ and PrIr$_2$Zn$_{20}$
above the transition temperatures of the antiferro- or ferro-quadrupole order and the superconductivity.
\cite{Onimaru, Matsubayashi}
\\

On the other hand, there remain some problems to be resolved.
Namely, $\lim_{T\to 0}C(T)\ne0$, which apparently breaks the Nernst law, and the anomaly of $\chi_m(T)$ is too small compared with that observed in experiments of 
PrV$_2$Al$_{20}$.
For the former problem, we have to take into account the existence of ordered states, which can release the entropy below their transition temperatures.
For the latter problem, as discussed in Sect. 7, we have to extend the model Hamiltonian to take into account the effects that are beyond the present model,
or consider the higher-order terms in the $1/N$-expansion as performed in the impurity version of the multichannel Anderson model discussed in Ref. \citen{Tsuruta2}.
Investigating various ordered states in the two-channel Anderson lattice model is also left for future study.

\begin{table}
\begin{tabular}{|l|c|c|}\hline
\hspace*{2em}$t\equiv T/E_0$ ($T_x/E_0\ll t\ll 1$)
& Single Channel & Two Channels\\ \hline
Electrical Resistivity & $a_1t^2$ & $a_2/(1+b/t)$\\ \hline
Chemical Potential & $\mu_0(1+a_3t^2)$ & $\mu_0(1+a_4\sqrt{t})$\\ \hline
Specific Heat & $\gamma T$ & $c_0(1-a_5\sqrt{t})$\\ \hline
Magnetic Susceptibility & $\chi_0(1-a_6t^2)$ & $\chi_0(1-a_7\sqrt{t})$\\ \hline
\end{tabular}
\caption{$T$ dependences of physical quantities in the case of single channel and two channels, in the region $T_x/E_0\ll t\ll 1$ with $t\equiv T/E_0$.}
\label{table}
\end{table}

\section*{Acknowledgments}
We are grateful to A. Sakai, S. Nakatsuji, and K. Matsubayashi, and T. Onimaru, K. Izawa, and Y. Machida  for stimulating discussions on the experimental results of Pr$A_2$Al$_{20}$ ($A=$V, Ti) and PrIr$_2$Zn$_{20}$, respectively.
In particular, we thank T. Onimaru for allowing us to use the experimental data of the resistivity in PrIr$_2$Zn$_{20}$ prior to publication.
This work is supported by
a Grant-in-Aid for Scientific Research on Innovative Areas
``Topological Quantum Phenomena'' (No. 22103003)
and by a Grant-in-Aid for Scientific Research (No. 25400369) 
from the Japan Society for the Promotion of Science.


\appendix
\section{Relationship among Self-Energies in Impurity Model, $d=\infty$ Lattice Model, and Lattice Model in Finite Dimensions}

\begin{figure}
\includegraphics[width=15cm]{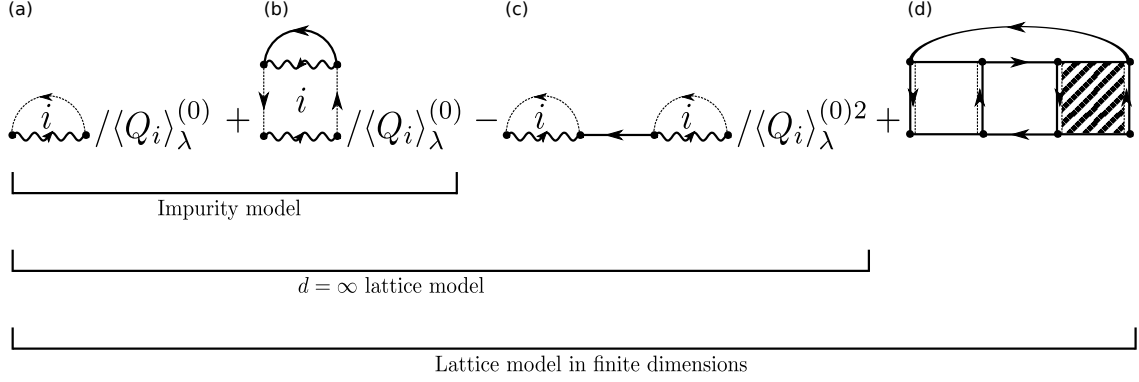}
\caption{Feynman diagram representation of the Dyson equations for the single-particle Green functions for the conduction electrons on the order of $O(1/N)$.}
\label{Fig:a1}
\end{figure}

In this appendix, we show the relationship among self-energies in the impurity model, $d=\infty$ lattice model, and lattice model in finite dimensions (treated by $1/N$-expansion formalism).
The self-energies of conduction electrons on the order of $O(1/N)$ are shown diagrammatically in Fig. \ref{Fig:a1}.
Diagrams in the impurity model are shown in Figs. \ref{Fig:a1}(a) and \ref{Fig:a1}(b),
from which the finite resistivity at $T=0$ is obtained.
However, the $T$ dependence of resistivity arising from these diagrams is negligibly weak at $T<T_{\rm K}$.
To obtain the $T^2$ dependence in the single-channel case and the $T^{1/2}$ dependence in the two-channel case, we have to further take into account the self-energies of pseudoparticles, shown by dotted and wavy lines, and vertex corrections.\cite{Tsuruta2}

In the $d=\infty$ lattice model, with the use of DMFT, we have to calculate the diagrams in Figs \ref{Fig:a1} (a)--\ref{Fig:a1}(c).
In the single-channel case, the resistivity given by Figs. \ref{Fig:a1}(a) and \ref{Fig:a1}(b) is cancelled by that given by Fig. \ref{Fig:a1}(c).
As in the case of the impurity, to obtain the correct $T$ dependence, we have to take into account the self-energies of pseudoparticles,
shown by dotted and wavy lines, and vertex corrections.
On the other hand, in the two-channel case, the cancellation is only partial, making the resistivity finite even at $T=0$.
This finite resistivity remains even if we take into account the self-energies of pseudoparticles,
shown by dotted and wavy lines, and vertex corrections.
Namely, the resistivity given by DMFT (e.g., Ref. \citen{Jarrell}) is proportional to (${\rm const.}-T$).

On the other hand,
in the lattice model in finite dimensions, we calculate all the diagrams Figs. \ref{Fig:a1}(a)--\ref{Fig:a1}(d) up to the order of $O(1/N)$.
In the single-channel case, 
the terms given by Figs. \ref{Fig:a1}(a)--\ref{Fig:a1}(c) are cancelled out,
and the Fermi-liquid-type resistivity proportional to $T^2$ is given by Fig. \ref{Fig:a1}(d).
In the multichannel case, however, the cancellation among Figs. \ref{Fig:a1}(a)--\ref{Fig:a1}(c) is not perfect so that we obtain the alternating series of $(T^*/T)^n$; thus, the resistivity becomes proportional to $1/(1+bT^*/T)$.
We have verified that this result remains valid even if we take into account the higher-order terms in $1/N$, which give an anomalous $T$ dependence of resistivity, i.e., $\propto -T^{1/2}$, in the case of the impurity model, although we do not show its explicit derivation here for conciseness of presentation.

\section{Formalism for Calculating the Specific Heat}

In this appendix, we show how 
the specific heat $C_V$ is calculated.

\begin{figure}
\includegraphics[width=10cm]{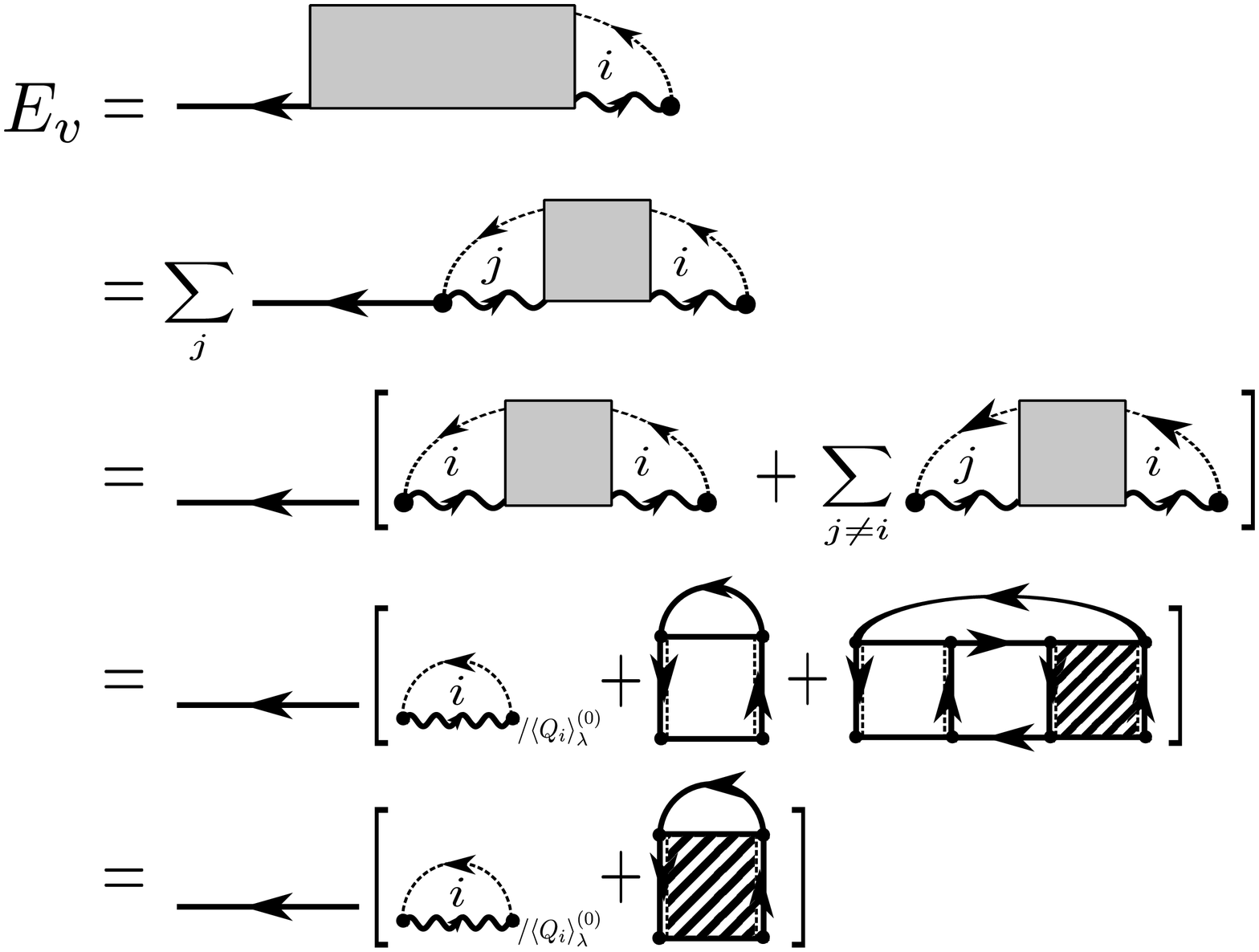}
\caption{
Feynman diagram representation for $E_v$.
The closed circle represents the hybridization $V$.
The solid line with an arrow represents the Green function of conduction electrons,
$G_{\mib{k}\tau\sigma}^{(1)}(i\epsilonn)$,
the dotted line with an arrow represents the Green function of pseudobosons,
$B_{i\tau}^{(0)}(i\nu_m)$,
the wavy line with an arrow represents the Green function of the slave fermions,
$F_{i\sigma}^{(1)}(-i\epsilonn+i\nu_m)$,
and the shaded square represents the vertex $\Gamma^{(0)}_{\mib{q}}$ given by Fig. 3(b).
}
\label{Fig:sh2}
\end{figure}

$C_V$ per site is given by
\beqa
C_V=\frac{1}{N_L}
\left(\frac{\partial \langle H\rangle}{\partial T}\right)_{V,N},\label{cc}
\eeqa
where $\langle H\rangle$ is the internal energy,
which is the average of Hamiltonian [Eq. (\ref{H})], with $\ep_{\mib{k}\tau\bar{\tau}}=0$,
\beqa
\langle H\rangle&=&E_c+E_f+E_v,
\eeqa
where
\beqa
E_c&=&\sum_{\mib{k},\tau,\sigma}\ep_{\mib{k}}\langle c^+_{\mib{k}\tau\sigma}c_{\mib{k}\tau\sigma}\rangle,\\
E_f&=&\sum_{i, \tau}\ep^{(0)}_{\Gamma_3}\langle b^+_{i\tau}b_{i\tau}\rangle
+\sum_{i, \sigma}\ep^{(0)}_{\Gamma_7}\langle f^+_{i\sigma}f_{i\sigma}\rangle,
\eeqa
and
\beqa
E_v&=&\frac{1}{\sqrt{N_L}}\sum_{i,{\mib k},\tau,\sigma}
\left (V \langle c^+_{\mib{k}\tau\bar{\sigma}} f^+_{i\sigma} b_{i \tau}\rangle
e^{-i \mib{k}\cdot\mib{R}_i} + {\rm h.c.} \right ).
\eeqa
With the use of the $1/N$-expansion, $E_c$ and $E_f$ are given by
\beqa
E_c&=&\sum_{\mib{k},\tau,\sigma}\ep_{\mib{k}}\int d\epsilon f(\epsilon)
\frac{-1}{\pi}{\rm Im}G^{(1)}_{\mib{k}\tau\sigma}(\epsilon+i0_+),
\eeqa
and
\beqa
E_f&=&\sum_{i}\ep^{(0)}_{\Gamma_3}n_{\rm pseudoboson}
+\sum_{i}\ep^{(0)}_{\Gamma_7}n_{\rm slave\mathchar`-fermion},
\eeqa
where the Green function $G^{(1)}_{\mib{k}\tau\sigma}(\epsilon+i0_+)$ is given by
Eq. (\ref{Eq:G1}), $n_{\rm pseudoboson}$ and $n_{\rm slave\mathchar`-fermion}$ are given below Eq. (\ref{Eq:nf2}).
The explicit Feyman diagrams for $E_v$ are shown in Fig. \ref{Fig:sh2}, and its analytic form is given as
\beqa
E_v&=&2\frac{1}{\sqrt{N_L}}\sum_{\sigma}\sum_{\tau}\sum_{i, \mib{k}}
V\langle c^+_{\mib{k}\tau\bar{\sigma}} f^+_{i\sigma} b_{i \tau}\rangle
e^{-i \mib{k}\cdot\mib{R}_i}
\non\\
&=&-2\frac{1}{\sqrt{N_L}}\sum_{\sigma}\sum_{\tau}\sum_{i, \mib{k}}V
\langle T_\tau f_{i\sigma}^+(-0_+)b_{i\tau}(-0_+)
c_{\mib{k}\tau\bar{\sigma}}^+(0)\rangle\e^{-i \mib{k}\cdot\mib{R}_i}\non\\
&=&-2\sum_{\sigma}\sum_{\tau}\sum_{\mib{k}}
T\sum_{\epsilonn}
e^{-i\epsilonn0_+}
G^{(1)}_{\mib{k}\tau\sigma}(i\epsilonn)
\Sigma_{\mib{k}\tau\sigma}(i\epsilonn)\non\\
&=&-2\sum_{\sigma}\sum_{\tau}\sum_{\mib{k}}
\int d\epsilon f(\epsilon)\frac{-1}{\pi}
{\rm Im}\left[G^{(1)}_{\mib{k}\tau\sigma}(\epsilon+i0_+)
\Sigma_{\mib{k}\tau\sigma}(\epsilon+i0_+)\right].
\eeqa

%

\section{Numerical Recipe}
In this appendix, we show a recipe for numerical calculations for
the integration with respect to the real frequency $\epsilon$, as in Eq. (\ref{Eq:E01}).
Explicitly, we use the trapezoidal rule:
\begin{eqnarray}
\int_{-\infty}^{\infty}d\epsilon A(\epsilon)\simeq\sum_{i=-L}^{L}\Delta\epsilon_iA(\epsilon_i).
\end{eqnarray}
To take finer meshes of summation in the low-frequency region,
we adopt a weighted mesh for summation over $\epsilon_i$ with the use of 
$\epsilon_i$ and $\Delta\epsilon_i$ given by
\begin{eqnarray}
\epsilon_i={\rm sgn}(i)\epsilon_1\frac{r^{|i|}-1}{r-1},
\end{eqnarray}
and
\begin{eqnarray}
\Delta\epsilon_i=\left\{
\begin{array}{ll}
\epsilon_1 & (i=0)\\
\frac{\epsilon_1}{2}r^{|i|-1}(r+1) & (i\ne0),\\
\end{array}
\right.
\end{eqnarray}
where ${\rm sgn}(x)$ is the sign function given by
\begin{eqnarray}
{\rm sgn}(x)\equiv\left\{
\begin{array}{ll}
1 & (x>0)\\
0 & (x=0)\\
-1 & (x<0).\\
\end{array}
\right. 
\end{eqnarray}
Note that the term with $\epsilon_0=0$ and $\Delta\epsilon_0=\epsilon_1$
is taken into account.
For explicit calculations, we adopt the parameters $L=100$, $\epsilon_1=10^{-3}D$, $\epsilon_L=2D$, and $r=1.04647534$, where $r$ is determined by
\beqa
\epsilon_L=\epsilon_1\frac{r^L-1}{r-1}.
\eeqa

\end{document}